\documentclass[12pt]{article}

\newenvironment{wileykeywords}{\textsf{Keywords:}\hspace{\stretch{1}}}{\hspace{\stretch{1}}\rule{1ex}{1ex}}

\usepackage{mhchem}
\usepackage{amsmath,amssymb,latexsym}
\usepackage{graphicx}% Include figure files
\usepackage{color}% Include colors for document elements
\usepackage{dcolumn}% Align table columns on decimal point
\usepackage{bm}% bold math
\usepackage[numbers,super,comma,sort&compress]{natbib}

\definecolor{background-color}{gray}{0.98}

\title{Quantum Computing in Pharma: A Multilayer Embedding Approach for Near Future Applications}
\author{R\'obert Izs\'ak\thanks{Riverlane Research Ltd, 1st Floor, St. Andrews House,
59 St. Andrews Street,
Cambridge,
CB2 3BZ}, 
Christoph Riplinger\thanks{FAccTs GmbH, K{\"o}ln, Germany},
Nick S. Blunt,$^*$
Bernardo de Souza,$^\dagger$
\\
Nicole Holzmann\thanks{Astex Pharmaceuticals, 436 Cambridge Science Park, Cambridge, CB4 0QA},$^*$
Ophelia Crawford,$^*$
Joan Camps,$^*$ 
\\
Frank Neese\thanks{Max-Planck Institut für Kohlenforschung 
Kaiser-Wilhelm-Platz 1
D-45470 Mülheim an der Ruhr 
Germany},
Patrick Schopf$^\ddagger$
}

\begin{document}

\maketitle

\begin{abstract}
Quantum computers are special purpose machines that are expected to be particularly useful in simulating strongly correlated chemical systems. The quantum computer excels at treating a moderate number of orbitals within an active space in a fully quantum mechanical manner. We present a quantum phase estimation calculation on F$_2$ in a (2,2) active space on Rigetti's Aspen-11 QPU. While this is a promising start, it also underlines the need for carefully selecting the orbital spaces treated by the quantum computer. In this work, a scheme for selecting such an active space automatically is described and simulated results obtained using both the quantum phase estimation (QPE) and variational quantum eigensolver (VQE) algorithms are presented and combined with a subtractive method to enable accurate description of the environment. The active occupied space is selected from orbitals localized on the chemically relevant fragment of the molecule, while the corresponding virtual space is chosen based on the magnitude of interactions with the occupied space calculated from perturbation theory. This protocol is then applied to two chemical systems of pharmaceutical relevance: the enzyme [Fe] hydrogenase and the photosenzitizer temoporfin. While the sizes of the active spaces currently amenable to a quantum computational treatment are not enough to demonstrate quantum advantage, the procedure outlined here is applicable to any active space size, including those that are outside the reach of classical computation.
\end{abstract}

\begin{wileykeywords}
quantum computers, enzymes, photochemistry, drugs, embedding.
%A list of five key words or phrases which best characterize the paper are required for indexing.
\end{wileykeywords}

\clearpage

%*****************Graphical Table of Contents******************** THIS IS MANDATORY *******************

\begin{figure}[h]
\centering
\colorbox{background-color}{
\fbox{
\begin{minipage}{1.0\textwidth}
\begin{center}
\includegraphics[height=100mm]{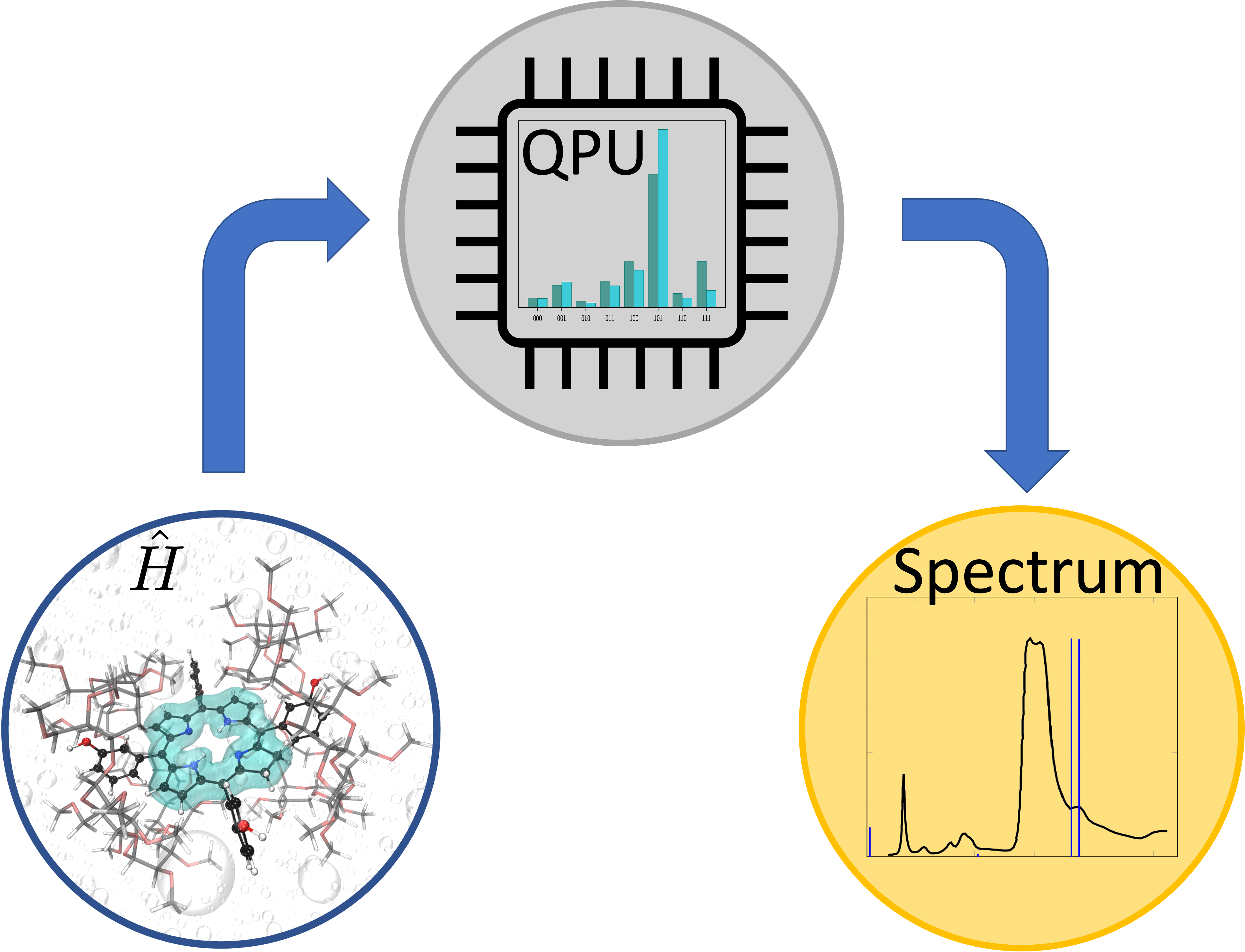}
\end{center}
% Pick only one of the two styles by uncommenting the corresponding \includegraphics
%\includegraphics[width=110mm,height=20mm]{cc.eps}
The active space selection scheme distributes the workload between classical and quantum computers, and forwards the necessary information to the quantum computer. The results are corrected for environmental effects via calculations on the classical computer and can be used to generate chemically relevant information, e.g., spectra.
\end{minipage}
}}
\end{figure}

% makes references listed with 1., 2., etc.  
  \makeatletter
  \renewcommand\@biblabel[1]{#1.}
  \makeatother

\bibliographystyle{apsrev}

\renewcommand{\baselinestretch}{1.5}
\normalsize

\clearpage

\section*{\sffamily \Large INTRODUCTION} 

While atoms are the fundamental building blocks of our world, it is chemical bonds in all their flavours and shapes that hold it together. And while thanks to physics we have a very good idea about how atoms behave, this little thing that is the chemical bond remains difficult to grasp. Gernot Frenking has dedicated his scientific life not only to the definition and classification of chemical bonds but to exploring those that are unusual and surprising and that open our minds up to new types of molecules and their potential applications. Our understanding of the chemical bond grew over the last decades as computational tools advanced. We are now at the advent of a new and powerful tool that we hope will help us in understanding bonds and their formation better: quantum computing. In the rest of this work, we will explore the near future applications of quantum computers in pharma and present an embedding protocol that will help optimize the workload on quantum computers.

Although the foundations of quantum mechanics were laid almost a century ago, it took decades before quantum chemical methods could be applied to sizeable chemical systems. The principal reason is that most methods scale badly with the number of orbitals and electrons involved in the system. One solution to the problem is to partition the system into subsystems that can be treated separately, and since the subsystems are smaller, the required calculations must also be cheaper. Any subsystem is then embedded in an embedding region consisting of the rest of the system. Embedding approaches can then be used to recover some property of the total system from its fragments (fragmentation) or to describe some feature in a selected central region at a higher level of theory (multilayer approaches). Over the years, several methods have been developed which differ in the precise nature of the partitioning. Separation based on the distinction between core and valence electrons goes back at least to the work of Fock and co-workers in 1940, \cite{fock1940incomplete} and foreshadows the frozen-core approximations still used today. Similar ideas were used by Lykos and Parr to give a separate treatment of sigma and pi electrons, \cite{lykos1956pi} and by McWeeny who developed a density matrix based approach for treating subsystems consisting of shells or groups of orbitals.\cite{mcweeny1959density} Philips and Kleinman justified the use of empirical core potentials in solid state calculations on a similar basis.\cite{phillips1959new} Summarizing and extending earlier work, Huzinaga and Cantu developed an effective Hamiltonian formulation that describes separated subsystems. \cite{huzinaga1971theory} Due to the present limitations on the number of orbitals that quantum computers can deal with, similar ways of limiting the orbital spaces visible to the quantum computer are unavoidable for chemical applications. Of particular interest in this respect is the concept of a complete active space (CAS)\cite{roos1980complete} that was introduced to tackle strongly correlated systems by introducing an active orbital space in which the exact full configuration interaction (FCI) expansion can be parametrized while the rest of the orbitals were treated as inactive (occupied) or external (virtual) orbitals. While this reduces the cost and produces qualitatively improved wavefunctions, it introduces a dependence on the choice of active spaces. Our goal then is to select an orbital subspace from the full set and to make this choice an optimal one. It should be noted that recently several automatic procedures have also been proposed that select an active space based on various criteria, including orbital entanglement,\cite{stein2016automated} atomic valance orbitals,\cite{sayfutyarova2017automated} first order perturbation theory\cite{khedkar2019active} and localization on a fragment.\cite{lau2021regional} While active space selection is certainly important for applications on the quantum computer, for quantitative uses, it is also important how the remaining parts of the chemical system are treated. In particular, the CAS expansion takes care of static electron correlation effects and not dynamic ones.\cite{knowles2000ab} While the CAS wavefunction may serve as a reference in multireference methods that recover dynamic correlation, and that might be carried out on the classical computer before or after the quantum computation, it may be difficult to obtain the necessary wavefunction information from the quantum computation. Next, we will discuss some embedding techniques that may also be employed for treating dynamical correlation effects in the environment and highlight their potential uses in quantum computing.

There is a great variety of embedding methods in use today. For a broad review, see Goez and Neugebauer.\cite{goez2018embedding} In a typical pharmaceutical calculation,\cite{lam2020applications} the environment is treated only at the molecular mechanics (MM) level, i.e., as an external potential or point charge field, while one or more central regions are described by quantum mechanics (QM). Such QM/MM methods fall in two categories. \cite{cao2018difference} Subtractive methods perform an MM calculation on the entire system and correct the result with the difference between a QM calculation and an MM calculation on the central region. Additive schemes contain an energy term for the central/QM region and one for the environmental/MM region, as well as a coupling term between the two. Another possible classification relies on the way the interaction between QM and MM layers is handled. \cite{bakowies1996hybrid} In mechanical embedding, only the MM region is influenced by point charges obtained in the QM region. In electrostatic embedding, the MM point charges are represented as an additional term in the Hamiltonian.  In polarizable  embedding, the MM point charges also change via the interaction with the QM density. As the latter approach is quite expensive, the most popular option currently is electrostatic embedding. Next, restricting our attention to QM/QM approaches still leaves a great variety of potential methods to be discussed. Instead of going over these in great detail, we refer the reader to the recent review of Sun and Chan,\cite{sun2016quantum} whose classification of QM/QM approaches into the broad categories of density functional embedding, Green's function embedding and density matrix embedding focuses attention to formalisms that originate in an exact description of the interaction between the subsystem and its environment. Density functional embedding\cite{cortona1991self,wesolowski1993frozen} is probably the most used approach in quantum chemistry, several reviews are available for the interested reader.\cite{jacob2014subsystem,libisch2014embedded,wesolowski2015frozen} Among the approaches that have been used in a wavefunction in density functional theory\cite{govind1998accurate} (DFT) context, the projector based approach of Manby and co-workers\cite{manby2012simple,lee2019projection} has been applied to sizeable chemical systems and it also has the advantage that the non-additive kinetic energy term vanishes due to the orthonormality of subsystem and environment orbitals.  Compared to Green's function and density matrix embedding, in density functional embedding, it remains difficult to ascertain whether the subsystem is entangled with its environment based on the subsystem density alone.  Green's function based approaches\cite{inglesfield1981method,georges1992hubbard,kotliar2006electronic} are more commonly used in condensed matter physics, although there have been several recent applications to quantum chemistry.\cite{lin2011dynamical,zgid2011dynamical,chibani2016self} In this category, dynamical mean field theory\cite{georges1992hubbard} (DMFT) maps the full many-body problem onto an impurity model and has been used with success to treat realistic systems. To avoid the practical and numerical difficulties associated with Green's functions,  density matrix embedding theory\cite{knizia2012density,knizia2013density,wouters2016practical} (DMET) uses the one-body reduced density matrix as the basic quantum variable. This method faces additional complications due to the fact that the high-level density matrices need not be idempotent, which have led to different cost functions and conditions which a total density assembled from fragments needs to satisfy. Since our goal is to describe a single central region accurately, rather than to assemble the whole system from fragments, the most interesting feature of DMET here is that it offers a way to construct bath orbitals (orbitals in the environment that are entangled with those in the system) that are exact at the mean field level and correspond to a set of occupied fragment orbitals spanned by intrinsic atomic orbitals (IAOs).\cite{knizia2013intrinsic} As it is possible to use DMET to define optimal link orbitals in a QM/MM calculation, \cite{sun2014exact} it should also be possible to use a similar approach in a QM/QM calculation, in which one of the QM regions is run on a quantum computer. Finally, it should be mentioned that embedding techniques may also be applied to excited state calculations, e.g., at the coupled cluster level.\cite{bennie2017pushing,parravicini2021embedded,izsak2020single}

Practical quantum advantage is defined here as a milestone for a quantum computer that consists of performing a useful task out of reach for a classical supercomputer. Quantum chemistry is one of the areas in which this milestone may be reached at an early stage.\cite{cao2019quantum,mcardle2020quantum} The expectation at the time of writing is that quantum phase estimation (QPE) is the most likely quantum algorithm to deliver quantum advantage in the chemistry domain.\cite{kitaev1995quantum,kitaev2002classical} As explained below, QPE can be used to, for example, calculate an FCI ground state energy with computational resources that are polynomial in the size of the chemical system; such a calculation would require exponential resources on a classical computer. Therefore, once sufficiently capable quantum computers are available, QPE could be used to calculate such energies for much larger chemical systems than with current supercomputers. However, as quantum resources will continue to remain limited, even once the milestone of practical quantum advantage is reached, embedding methods will be important to allow calculations to be performed on a range of systems. Thus, an optimal leverage of the capabilities of QPE requires an optimal choice of active space.

Running QPE for quantum-advantage-sized active spaces requires `deep' quantum circuits -- circuits with many quantum operations.  To run such deep circuits reliably one needs `quantum error correction': a procedure to effectively reduce the noise on the quantum computer at a cost of an overhead in the number of qubits. Although the field has already seen demonstrations of quantum error correction,\cite{chen_exponential_2021} currently available quantum computers are not yet quantum error corrected. For this reason, hybrid quantum-classical approaches are commonly used for early demonstrations of the applicability of quantum computers.  Hybrid algorithms are distributed between classical and quantum computers. Their quantum parts involve shallow quantum circuits, that are affected by noise to a lesser degree.

The variational quantum eigensolver (VQE)\cite{peruzzo2014variational,mcclean2016theory} is such a hybrid algorithm in the chemistry domain. VQE was developed to solve a variational optimization problem for which the Hamiltonian matrix elements are calculated on a classical computer. It is mostly used to solve the unitary coupled cluster\cite{taube2006new} (UCC) problem for chemical systems. The choice of UCC as opposed to traditional coupled cluster methods used on classical computers is due to the fact that operations on qubits are carried out as unitary transformations. The VQE-UCC approach also differs from classical coupled cluster theory in that it is variational, although it is still usually truncated at the singles and doubles (UCCSD) level. Due to long runtimes associated with the quantum and classical elements of this optimisation, this results in a computational problem that is only tractable for relatively small systems using small basis sets. Its applicability can be extended\cite{barkoutsos2018quantum,romero2018strategies} by freezing orbitals, reducing the full orbital space to a smaller active space or by discarding UCC amplitudes for which the first order estimate is small. Active space selection can also be optimized by transforming into a natural orbital basis calculated at the M{\o}ller-Plesset (MP2) level. \cite{verma2021scaling} Among embedding approaches, DMFT\cite{Bauer2016} and DMET\cite{rubin2016hybrid} implementations on quantum computers have already been described and the cost of DMET calculations has already been estimated. \cite{yamazaki2018towards} Energy-weighted DMET\cite{tilly2021reduced} and Gutzwiller variational embedding\cite{Yao2021} approaches have been tested on current quantum processors. For other choices of ans\"atze and more recent developments, see a recent survey.\cite{fedorov2022vqe}

In the remainder of this paper, we will propose an embedding method based on fragment based layering and natural orbital approaches. We will show how this method can be used to select a central fragment in large chemical systems, then proceed to treat the central fragment on a quantum computer and the environment on a classical one. While the example systems we present are limited by the current capacity of available quantum computers, the basic formulation does not depend on the active space size. Thus, as more powerful quantum computers become available, only the active space parameters need change in our approach.

\section*{\sffamily \Large THEORY}

\section*{\sffamily \Large The Structure of the Hamiltonian}

The molecular Hamiltonian can be written in general as
\begin{equation}
\hat{H} = E_0 + h^q_p E^p_q + \frac{1}{2}g^{qr}_{ps} E^{ps}_{qr},
\label{Ham}
\end{equation}
with $E_0$ being the nuclear repulsion term, $h^q_p$ the sum of the kinetic energy operators of the electrons and the nuclear-electronic attraction terms, and $g^{qr}_{ps}$ the two-electron integrals representing interelectronic repulsion. By the Einstein convention, repeated indices are summed up. The excitation operators $E^p_q$ are the generators of the unitary group and they are simply given as
\begin{equation}
E^p_q = a^{p\alpha}a_{q\alpha} + a^{p\beta}a_{q\beta},
\label{Egen}
\end{equation}
with $a^{p\alpha}$ being the creator of the spin-orbital $p\alpha$ and $a_{q\alpha}$ the annihilator of $q\alpha$. Summing over the spin labels yields ($\alpha$, $\beta$) a spin-free formulation and it is enough to refer to the spatial labels $p,q,\ldots$ instead. In quantum computing, the Hamiltonian is usually expressed in the spin-orbital basis, but the relevant expressions are easily retained by substituting Eq.~\eqref{Egen} into Eq.~\eqref{Ham} and remembering that $E^{ps}_{qr} = E^{p}_{q}E^{s}_{r}-\delta^{s}_{q}E^{p}_{r}$. 

On the quantum computer, the fermion operators $a^{p\alpha}$ and $a_{q\alpha}$ are transformed into qubit operators\cite{tranter2018comparison} and the complexity of the calculation depends on the magnitude of the 1-norm of this qubit Hamiltonian and through that on contributions from large integrals $h^q_p$ and $g^{qr}_{ps}$. Since Eq.~\eqref{Ham} is invariant under unitary transformations of orbitals, this property may be exploited to reduce the number of large terms in the expression and/or reduce the 1-norm using standard localization techniques.\cite{koridon2021orbital} For a normalized wavefunction $\Psi$, the energy is evaluated as the expectation value
\begin{equation}
E = \langle\Psi|\hat{H}|\Psi\rangle =
E_0 + h^q_p \Gamma^p_q + \frac{1}{2}g^{qr}_{ps} \Gamma^{ps}_{qr},
\label{Efun}
\end{equation}
in which $\Gamma^p_q = \langle\Psi|E^p_q|\Psi\rangle$, $\Gamma^{ps}_{qr} = \langle\Psi|E^{ps}_{qr}|\Psi\rangle$ are the spin-free one- and two-body reduced density matrices (RDM). Natural orbitals ($\{\tilde{p}\}$) diagonalize $\Gamma^p_q$, i.e.,  
\begin{equation}
\Gamma^{\tilde{p}}_{\tilde{q}} = n_{\tilde{p}}\delta^{\tilde{p}}_{\tilde{q}}.
\end{equation}
In this basis, only those diagonal terms $h^{\tilde{p}}_{\tilde{p}}$ need to be evaluated for which the occupation number $n_{\tilde{p}}$ is larger than a given threshold in order to evaluate the one-body term to a given accuracy. The structure of the much more numerous two-body terms also changes in the natural orbital basis. Consider the relationship
\begin{equation}
\Gamma^{ps}_{qr} = \Gamma^{p}_{q}\Gamma^{s}_{r} - \frac{1}{2}\Gamma^{s}_{q}\Gamma^{p}_{r} + \Lambda^{ps}_{qr}.
\end{equation}
For independent particle approaches, the two-body cumulant\cite{kutzelnigg1997normal} $\Lambda^{ps}_{qr}$ is zero, and the 2-RDM is also pairwise diagonal (giving rise to a direct and an exchange term). Unfortunately, $\Lambda^{ps}_{qr}$ is not negligible in general, although it should be noted that there are efficient pair natural orbital approaches that circumvent this problem by diagonalizing certain blocks of $\Gamma^{ps}_{qr}$. \cite{kutzelnigg1963loesung} Furthermore, the densities are not available before the calculation, and practical approaches must start with a density estimate that can be used for approximate diagonalization.\cite{neese2009efficient} Nevertheless, natural orbitals may still be used to rank the orbitals based on the occupation numbers. Alternatively, or sometimes in combination with (pair) natural orbital approaches, orbital localization techniques may also be used.\cite{neese2009efficient} The latter involve a unitary transformation that confines most of the orbitals to a relatively small region within the molecule. This means that the number of large $h^q_p$ and $g^{qr}_{ps}$ is reduced since distant orbitals essentially do not overlap. As will be seen later, both these approaches can be used to make a computation feasible on a quantum computer.

\section*{\sffamily \Large Quantum Algorithms and The Wavefunction Ansatz}

There exist two main algorithms for performing chemical calculations on a quantum computer -- VQE and QPE. In VQE, a functional of the form Eq.~\eqref{Efun} is minimised for some wavefunction $\Psi$ under the constraint that $\Psi$ is normalized. It is a hybrid approach that requires an ansatz -- a parametrization of the wavefunction. Inspired by classical approaches, and the requirement that operations on qubits be unitary, the ansatz of choice is the unitary coupled cluster singles and doubles (UCCSD) approach, defined with respect to the Hartree-Fock reference state $|0\rangle$ as
\begin{equation}
|\Psi\rangle = e^{\hat{T}-\hat{T}^\dagger}|0\rangle.
\end{equation}
The cluster operator $\hat{T}$ is truncated at the singles and doubles level,
\begin{equation}
\hat{T} = t^{i}_{a}E^a_i + \frac{1}{2}t^{ij}_{ab}E^a_iE^b_j,
\end{equation}
where again, spin-orbital expressions can be obtained by substituting Eq.~\eqref{Egen}. The indices $i,j,\ldots$ refer to orbitals occupied in the HF reference, while $a,b,\ldots$ denote virtual ones, as opposed to the general labels $p,q,\ldots$ that may refer to any orbital. As in the classical case, the result of the calculation is the energy value and the wavefunction parameters (cluster amplitudes) $t^{i}_{a}$, $t^{ij}_{ab}$. The UCCSD approach yields a Hermitian eigenvalue problem that can be variationally solved, which is less of a problem than on classical computers and removes some of the problems associated with the traditional solution (bond dissociation catastrophy).\cite{taube2006new,knowles2000ab}

In contrast to VQE, the QPE algorithm finds eigenvalues of a unitary operator $\hat{U}$ directly, rather than ansatz-constrained approximations to these eigenvalues. As the absolute value of the eigenvalues of a unitary operator is 1, the algorithm yields a phase $\theta_k$ for an eigenstate $|\Psi_k\rangle$ of the unitary operator, 
\begin{equation}
\hat{U}|\Psi_k\rangle = e^{2\pi i\theta_k}|\Psi_k\rangle.
\end{equation}
To relate this to eigenvalues of the Hamiltonian in Eq.~\eqref{Ham}, one chooses a unitary $\hat{U}$ related to the Hamiltonian $\hat{H}$, e.g., the propagator
\begin{equation}
\hat{U} = e^{-i\hat{H}t},
\end{equation}
and the eigenvalues of $\hat{H}$ can then be simply calculated from the phases $\theta_k$.

Since the eigenstates $|\Psi_k\rangle$ cannot be prepared directly, one applies the QPE algorithm to an initial state with a large overlap probability with the target eigenstate $|\Psi_k\rangle$ -- e.g., the Hartree-Fock state $|0\rangle$ if the target is the ground state -- and the algorithm succeeds in finding the eigenvalue $\theta_k$ with that probability.

A good implementation of QPE requires a good implementation of the unitary $\hat{U}$ on the quantum computer. The area of research associated with finding an optimal implementation is called quantum simulation. Several quantum simulation methods are available. Below we use the Trotter approximation,\cite{trotter1959product} that breaks the time interval $t$ into short-duration intervals.

The QPE algorithm presents a number of advantages over VQE, including that it does not require an explicit parametrization of an ansatz.  Rather, when the algorithm terminates, the exact eigenstate $|\Psi_k\rangle$ is stored in quantum memory, where $N$ qubits are used to represent a state of (roughly) $N$ spin orbitals. Thus, the exponential scaling of the FCI problem can be avoided while calculating the energy. It remains a question how to extract further information out of the exact wavefunction in quantum memory, and the algorithm is also less feasible on currently available quantum computers, because the required circuits have a much higher depth.

\section*{\sffamily \Large Embedding Approaches}

Fidelities of current quantum computers are not sufficiently high to perform VQE or QPE with large numbers of qubits, as the required quantum circuits quickly become too long. Instead, various embedding approaches may be used to make chemical computations more manageable on a quantum computer, albeit still insufficient for practical quantum advantage. Broadly defined, embedding approaches partition the system into two or more subsystems and then either treat these at different levels of theory (our main focus here) or attempt to assemble a property of the total system from those of its parts. As in the classical case, complete active space (CAS) approaches are a case in point: the FCI or UCCSD problem is solved in a selected set of active orbitals (A) and the environment (B) is treated at the HF level. Note that orbitals are rarely optimized on the quantum computer, thus this treatment corresponds to a CASCI rather than a CASSCF calculation. \cite{tilly2021reduced} In any case, the relevant Hamiltonian can be obtained by a projection to the active space ($\hat{P}_A$),
\begin{equation}
\hat{P}_A\hat{H}\hat{P}_A = 
E_B + f(\mathbf{D}_B)^u_t E^t_u + \frac{1}{2}g^{uv}_{tw} E^{tw}_{uv},
\end{equation}
where $E_B$ is the HF energy calculated from orbitals in B, and $f(D_B)$ is the Fock matrix calculated calculated from the HF density of B, $\mathbf{D}_B$. Formally, this Hamiltonian is very similar to Eq.~\eqref{Ham}, but the set of active occupied and virtual orbitals labelled as $t,u,\ldots$ is much smaller than the full set containing $p,q,\ldots$.

In multilayer embedding approaches commonly used in pharmaceutical calculations, the central layer is treated at a higher level than the environment. Nevertheless, the treatment of the environment need not be restricted to the HF level, as the discussion so far might suggest. Iterative embedding approaches may be set up in a way that the central region interacts with the environment in a macro-loop. However, a simpler approach might suffice for many purposes: in subtractive embedding schemes, the total energy of the system $E_{\text{total}}(\text{A}+\text{B})$ can be calculated as
\begin{equation}
E_{\text{total}}(\text{A}+\text{B}) = 
E_{\text{low}}(\text{A}+\text{B}) + E_{\text{high}}(\text{A}) - E_{\text{low}}(\text{A}),
\label{EMBsub}
\end{equation}
where a low level calculation on the whole system $E_{\text{low}}(\text{A}+\text{B})$ is corrected by the difference between a higher level calculation on A ($E_{\text{high}}(\text{A})$) and a low level calculation on A ($E_{\text{low}}(\text{A})$). In the present context, the higher level may be VQE-UCCSD or QPE, while the low level approach could be some classical correlation method, e.g., CCSD. This still leaves us with the problem of selecting an active space, which will be investigated next.

\section*{\sffamily \Large Orbital Selection in the Occupied Space}

As local orbitals are confined to a specific region in the molecule, it makes sense to use them in the common situation when it is known that a reaction affects a certain region more than others. The idea behind the intrinsic atomic orbital (IAO) approach\cite{knizia2013intrinsic} is that due to the high variational freedom in large basis set expansions, the degree of delocalization of canonical molecular orbitals (MOs) is larger compared to an expansion in a minimal basis consisting of free-atom AOs. On the other hand, polarization functions of large bases are necessary in order to properly represent the molecular environment. Thus, the occupied MOs are spanned in a second minimal basis set, which is constructed so that this subspace is exactly spanned.\cite{knizia2013intrinsic,lu2004molecule} For a start, some commonly used minimal basis set is selected and the MO coefficients \textbf{C} are projected to this minimal basis and back to the original one,
\begin{equation}
\tilde{\mathbf{C}} = \text{orth}\left(\mathbf{P}_{12}\mathbf{P}_{21}\mathbf{C}\right),
\end{equation}
where for any $\mathbf{X}$
\begin{equation}
\text{orth}(\mathbf{X}) = \mathbf{X}(\mathbf{X}^T\mathbf{S}_1\mathbf{X})^{-1/2}
\end{equation}
denotes Löwdin's symmetric orthonormalization. Here, $\mathbf{S}_1$ is the overlap associated with the original basis, $\mathbf{S}_2$ with the minimal basis and $\mathbf{S}_{12}=\mathbf{S}_{21}^T$ is the mixed overlap between the two sets, $\mathbf{P}_{12}=\mathbf{S}_1^{-1}\mathbf{S}_{12}$ and $\mathbf{P}_{21}=\mathbf{S}_2^{-1}\mathbf{S}_{21}$. The new set of MO coefficients $\tilde{\mathbf{C}}$ are orthonormal and they are spanned in a minimal basis set. These ``depolarized'' coefficients are then used to construct the IAO minimal basis with the transformation matrix
\begin{equation}
\mathbf{A} = \text{orth}\left(\mathbf{CC}^T\mathbf{S}_1\tilde{\mathbf{C}}\tilde{\mathbf{C}}^T\mathbf{S}_1\mathbf{P}_{12} + 
(\mathbf{I}-\mathbf{CC}^T\mathbf{S}_1)(\mathbf{I}-\tilde{\mathbf{C}}\tilde{\mathbf{C}}^T\mathbf{S}_1)\mathbf{P}_{12}\right),
\end{equation}
The first term transforms the minimal basis into the depolarized occupied space and then spans the polarized occupied space in terms of these, the second term does the same with the virtual orbitals.\cite{knizia2013intrinsic,lu2004molecule} The new orthonormal IAO basis $\{|\rho\rangle\}$ is thus given by
\begin{equation}
|\rho\rangle = \sum_{\mu} A_{\mu\rho}|\mu\rangle,
\end{equation}
where $\{|\mu\rangle\}$ is the original orbital basis. Finally, the occupied orbital coefficients in the IAO basis are given by the inverse transformation
\begin{equation}
\bar{\mathbf{C}} = \mathbf{A}^T\mathbf{S}_1\mathbf{C}.
\label{CIAO}
\end{equation}

As originally proposed, the construction of IAOs is followed by a Pipek-Mezey localization\cite{pipek1989fast} step to yield intrinsic bond orbitals (IBOs). Here, we start with some localization process and use the IAO transformation in Eq.~\eqref{CIAO} to span the result in a basis that can be easily assigned to a fragment. Suppose now that a system is partitioned into a central region A and an environment B, and that a Hartree-Fock calculation is available for the \emph{entire} system (A+B). If $\bar{\mathbf{C}}_i$ contains the MO coefficients for orbital $i$ in terms of IAOs, these may be projected to fragment A using the projector $\mathbf{P}_A$ and their norm $Q_i$ may be evaluated as
\begin{equation}
Q_i = \bar{\mathbf{C}}_i^T\textbf{P}_A\bar{\mathbf{C}}_i.
\end{equation}
If this norm is larger than a certain threshold, we may conclude that the orbital is mostly localized on fragment A. Since the (unprojected) MOs are normalized, this threshold should not be much smaller than 1.  In a large molecule with a small central fragment, this will reduce the list of relevant orbitals significantly. Let us label these $i_A,j_A,\ldots$ and refer to them as fragment occupied orbitals. We may still want to select only some of these. To estimate their significance, we may rely on perturbation theory to calculate the approximate natural orbitals from first order semi-canonical\cite{neese2009efficient} M{\o}ller-Plesset amplitudes $t^{i_A j_A}_{ab}$,
\begin{equation}
t^{i_A j_A}_{ab} = -\frac{g^{i_A j_A}_{ab}}{\varepsilon_a + \varepsilon_b - f_{i_A i_A} - f_{j_A j_A}}
\end{equation}
constituting the occupied block of the density, $D_{i_A j_A}$. Here, $\varepsilon_a$ denotes the orbital energy of the virtual orbital $a$, while $f_{i_A i_A}$ is the diagonal element of the Fock matrix corresponding to the fragment orbital $i_A$. Since there is a large number of virtual pairs, the problem may be simplified by restricting the relevant summation to diagonal pairs in the M{\o}ller-Plesset density
\begin{equation}
D_{i_A j_A} = 2\sum_{k_A a}\left(t^{i_A k_A}_{aa}t^{k_A j_A}_{aa}+t^{k_A i_A}_{aa}t^{j_A k_A}_{aa}\right).
\end{equation}
This density may also be regarded as an averaged orbital specific density for occupied orbitals, in a similar sense as it was defined for virtuals in the literature.\cite{yang2011tensor} Diagonalizing it yields
\begin{equation}
D_{i_A j_A} = \sum_{\tilde{i}_A} d_{i_A\tilde{i}_A}n_{\tilde{i}_A}d_{i_A\tilde{j}_A},
\end{equation}
and the occupation number $n_{\tilde{i}_A}$ may be used to select active occupied orbitals $\tilde{i}_A,\tilde{j}_A,\ldots$. The transformation matrix $d_{i_A\tilde{i}_A}$ can be used to change to this new basis. The approach may be simplified even further by calculating the diagonal elements $D_{i_A i_A}$ only, as an approximation to $n_{\tilde{i}_A}$. 

So far, our approach shows similarity to the fragment based active space selection approach as far as selecting the active occupied space is concerned,\cite{lau2021regional} except for the optional perturbative selection scheme introduced here. The use of IAOs is also a feature shared by many embedding approaches, including DMET\cite{knizia2013density} and projector-based embedding.\cite{manby2012simple} Formally, DMET uses a similar active space Hamiltonian, but it is mostly used as a fragmentation approach, as opposed to the multilayer scheme proposed here. While DMET bath orbitals were also proposed as optimal link orbitals in multilayer QM/MM schemes,\cite{sun2014exact} our implementation does not make use of this approach at the moment. There is formal similarity to projector based embedding as well, except that we do not make use of DFT for defining an embedding potential.

\section*{\sffamily \Large Orbital Selection in the Virtual Space}

Once the occupied orbitals have been selected, the M{\o}ller-Plesset one-body density (averaged pair density) may be constructed for the pairs $\tilde{i}_A \tilde{j}_A$ within the active space
\begin{equation}
D_{ab} = 2\sum_{\tilde{i}_A\tilde{j}_A c}\left(\tilde{t}^{\tilde{i}_A\tilde{j}_A}_{ac}t^{\tilde{i}_A\tilde{j}_A}_{bc}+\tilde{t}^{\tilde{i}_A\tilde{j}_A}_{ca}t^{\tilde{i}_A\tilde{j}_A}_{cb}\right)
\end{equation}
from the first order amplitudes $t^{\bar{i}_A\bar{j}_A}_{b^{\prime}c^{\prime}}$ and contravariant amplitudes $\tilde{t}^{\bar{i}_A\bar{j}_A}_{b^{\prime}c^{\prime}}=2t^{\bar{i}_A\bar{j}_A}_{b^{\prime}c^{\prime}}-t^{\bar{i}_A\bar{j}_A}_{c^{\prime}b^{\prime}}$, and diagonalized
\begin{equation}
D_{ab} = \sum_{\tilde{a}} d_{a\tilde{a}}n_{\tilde{a}}d_{b\tilde{a}}.
\end{equation}
The significant virtuals can then be selected based on a threshold imposed on the natural orbital occupation number $n_{\tilde{a}}$ and the surviving natural orbitals can be obtained simply via the transformation matrix $d_{a\tilde{a}}$. It is important to emphasize that only the active occupied orbitals are selected based on the criterion that they are localized on the selected fragment. The virtuals are selected based on the interaction with the already selected occupied space and may have significant contributions from basis functions outside the fragment.

This part of our selection scheme resembles the perturbative active space selection scheme proposed earlier,\cite{khedkar2019active} except that we use M{\o}ller-Plesset amplitudes in our scheme. Thus, our approach uses established active space selection techniques at each stage and combines them in a way that is best suited for the given orbital space: the occupied space is selected based on the local changes induced by chemical processes, while the virtual space is chosen by the criterion of strongest perturbative interaction with the selected occupied space. Once the active space selection has been carried out, the outer layers are treated using a subtractive embedding approach outlined above.

\section*{\sffamily \Large COMPUTATIONAL DETAILS}

QPE calculations on F$_2$ (2,2) were performed using Rigetti's Aspen-11 quantum processor (QPU). The Aspen-11 QPU is a superconducting device with 40 qubits and both 2-fold and 3-fold connectivity available. Calculations were submitted to Aspen-11 using Rigetti's Quantum Cloud Services (QCS) platform\cite{Karalekas_2020} and software stack.\cite{smith2016practical} This includes the pyQuil library to implement the QPE circuit and quilc to compile it to native Aspen-11 gates.

All chemistry calculations on the classical computer were carried out with a development version of the ORCA program package~\cite{ORCA}. ORCA generates the integral files that are required by calculations on or simulations of the quantum computer either in the FCIDUMP format\cite{knowles1989determinant} or as a Jordan-Wigner qubit Hamiltonian\cite{jordan1928ueber,tranter2018comparison} printed in a format recognizable to the QubitOperator class in OpenFermion.\cite{mcclean2020openfermion}

We used Riverlane-developed software to perform emulated quantum computational calculations. The software enables resources to be estimated, circuits to be constructed and emulated computations to be performed for both VQE and QPE calculations with a range of inputs. We also made use of OpenFermion~\cite{mcclean2020openfermion} in order to form the Hamiltonian and parts of the circuits to be emulated. For the emulation of the computations, we used ProjectQ.\cite{steiger2018projectq,haner2018software}

The CC bond length in dimethyl-acetylene was scanned between 0.85 {\AA} and 2.85 {\AA} with a 0.05 {\AA} increment and using the orbital coefficients from the previous step as an initial guess. For the VQE calculation, the (6,6) active space integrals were generated and used as input for the Riverlane-developed software. The qubit Hamiltonian was constructed using the Jordan-Wigner transformation.\cite{jordan1928ueber} A 10$^{-6}$ Hartree accuracy was requested in simulations using ProjectQ.\cite{steiger2018projectq,haner2018software}

The initial protein structure and force field file was kindly provided by the Reiher group.\cite{finkelmann2014hydrogen} In the reaction path calculations and subsequent single point calculations of the [Fe] hydrogenase system, the additive QM/MM method with electronic embedding was applied.  The QM region includes the FeGP cofactor up to the phosphate linker, the side chain of Cys175 coordinating to Fe of FeGP, the pterine, imidazoline and phenyl part of the substrate, as well as the hydrogens that are reacting with the substrate.  The active region is built around the Fe center. It includes all atoms that have a distance of less than 5 {\AA} to the iron center of FeGP, plus the backbone or sidechain (for proteic residues) or full molecules (for non-proteic groups) that these atoms belong to. These were the full molecules or residues FeGP, H4MPT, waters WAT1344 and WAT1070, the backbone atoms of PRO202 and VAL205, and the side chain atoms of TRP148, CYS176, HIS201 and VAL205. In the reaction pathway calculations, the TPSS functional\cite{tao2003climbing} was used with the D3 dispersion correction\cite{grimme2010consistent}  and the def2-TZVP basis set\cite{weigend2005balanced} for the QM part, and the AMBER force field\cite{wang2004development,duan2003point,lee2004distinguish} (after conversion to the prms format as required by the ORCA software using the orca\_mm module) was used for the MM part of the calculation. Covalent bonds at the boundary between QM and MM subsystem are capped in the QM subsystem calculation using the hydrogen link atom approach. The atomic charges of the MM atoms at the boundary are shifted using the charge shift scheme in the QM subsystem calculation in order to avoid overpolarization of the QM subsystem.\cite{lin2005redistributed} Minimum energy paths and transition states for the enzymatically catalyzed reaction were obtained on the QM/MM level using the NEB-CI and NEB-TS methods, as implemented in ORCA, starting from the optimized intermediate structures.\cite{asgeirsson2021nudged} Subsequent single point calculations on the extrema were carried out at the DLPNO-CCSD (domain-based local pair natural orbital CCSD)  method,\cite{riplinger2013efficient,pinski2015sparse,riplinger2016sparse} with the def2-TZVPP basis set, \cite{weigend2005balanced} as the method of choice in the QM region. 

The VQE calculation within the (4,4) active space made use of ORCA integrals used by the Riverlane-developed software, as in the case of dimethyl-acetylene. The qubit Hamiltonian was constructed using the Jordan-Wigner transformation\cite{jordan1928ueber}  and an accuracy of 10$^{-6}$ Hartree was requested.

The geometry of temoporfin was taken from the last snapshot of a QM/MD calculation kindly provided by the Monari and Catak groups. \cite{aslanoglu2020optical} Our calculations made use of the implementation of the ONIOM3 method in ORCA, in which the model subsystem (temoporfin) and the intermediate subsystem (two cyclodextrin carriers) are treated at the QM level (QM1 and QM2), and the environment (water) is treated at the MM level. \cite{lundberg2009oniom} The energy is evaluated using the mixed QM1/QM2/MM method as proposed by Chung et al., combining the subtractive ONIOM2 method together with the additive QM-MM method,\cite{chung2010modeling,chung2012oniom} 
\begin{equation}
E_{\text{ONIOM3}}=E_{\text{QM1,model}}+E_{\text{QM2,intermediate}}-E_{\text{QM2,model}}+ E_{\text{MM}}+E_{\text{QM-MM}}
\end{equation}
The electrostatic embedding scheme is used, where the intermediate subsystem (temoporfin plus organic carrier) calculation is polarized by the MM atomic charges, and the model subsystem (temoporfin) is polarized by the MM atomic charges and the Hirshfeld charges of the organic carrier atoms that are obtained from the intermediate subsystem calculation at the QM2 level.\cite{mayhall2010oniom} Here, QM2 is chosen to be the HF-3c model.\cite{sure2013corrected} The system was optimized on the ONIOM3 level with the B3LYP functional\cite{stephens1994ab} with the def2-TZVP basis set\cite{weigend2005balanced} as the QM1 method. In this optimization only the temoporfin molecule was active, i.e., the organic carrier and surrounding waters were kept fixed. The D3 dispersion correction as parametrized by Grimme and co-workers was also applied. \cite{grimme2010consistent} The B3LYP functional has already been used with success to obtain geometries in our earlier studies on other families of dyes. \cite{berraud-pache2020unveiling,lechner2021excited} The TightSCF keyword was used to set convergence criteria for energy calculations (thresholds in atomic units: $10^{-8}$ for energy and $10^{-5}$ for the orbital gradient). Vertical excitation energies and transition dipoles were computed with the DLPNO variant of the similarity transformed equation of motion (STEOM) method, DLPNO-STEOM-CCSD,\cite{nooijen1997similarity,dutta2016towards} as QM1 on the ONIOM3 geometries using the def2-TZVPP basis set\cite{weigend2005balanced} and the matching auxiliary basis set.\cite{hellweg2007optimized} For the purposes of calculating vertical excitation energies, the four hydroxyphenyl groups were excluded from the model subsystem calculation at the STEOM-DLPNO-CCSD (QM1) level and were instead included in the intermediate subsystem calculations at the HF-3c level (QM2). The atomic charges of the QM2 atoms at the QM1-QM2 boundary are shifted using the charge shift scheme in the QM1 subsystem calculation in order to avoid overpolarization of the QM1 subsystem.\cite{lin2005redistributed} The RIJCOSX \cite{neese2009efficient_cosx,izsak2011overlap} approximation was used in the STEOM integral dressing step. Ten roots were requested using ORCA's `TightPNO' settings~\cite{liakos2015exploring}, i.e., setting the following thresholds: T$_\text{CutPairs} = 10^{-5}$, T$_\text{CutDO} = 5\times 10^{-3}$, T$_\text{CutPNO} = 10^{-7}$, T$_\text{CutMKN} = 10^{-3}$. Since the quality of the singles (diagonal) PNOs is especially important,  these are generated as a separate set using a tighter than usual threshold of T$_\text{CutPNOSingles} = 10^{-11}$. The cutoff values for the natural orbital occupation numbers in the active space selection procedure for the occupied and virtual orbitals (`Othresh' and `Vthresh', respectively) were set to $5.0\times 10^{-3}$.\cite{dutta2017automatic} 

Using integrals from ORCA, the QPE calculations for the (4,4) active space of temoporfin were carried out using Riverlane-developed software together with ProjectQ~\cite{steiger2018projectq,haner2018software} to perform the simulation. The occupied orbitals were localized using the Pipek-Mezey method\cite{pipek1989fast} for the active space selection scheme. The Bravyi-Kitaev transformation\cite{Bravyi2002, Seeley2012} was used to obtain the qubit form of the Hamiltonian. We use the iterative QPE algorithm\cite{iQPE} and first-order Trotterisation with four Trotter steps. We perform simulations for the ground and first four singlet excited states. For each, we measure to 10 bits of accuracy, which requires 1023 controlled-unitary applications. The data qubits are initialised from a single determinant, chosen to have a large overlap with each respective state. We perform between 1 and 10 shots in order to obtain the desired state from each QPE experiment.

Threshold values for judging whether deviations are significant are assumed to be $\sim$1 kJ/mol for calculating energy differences along the minimal energy pathway. For vertical excitation energies, it is desirable that the computational method be able to reproduce the positions of band maxima within about 0.1~eV. Our earlier experience shows that the DLPNO-STEOM method can deliver such accuracy for various dyes.\cite{berraud-pache2020unveiling,lechner2021excited}

\section*{\sffamily \Large RESULTS AND DISCUSSION}

\section*{\sffamily \Large Performing QPE on a quantum processor}

We first investigate a simple example of QPE on a real quauntum processor. In particular, we use Rigetti's Aspen-11 QPU. This is a superconducting device with 40 qubits with 2 and 3-fold qubit connectivity. Further details about the QPU and Rigetti's software layer were given in the ``Computational Details'' section.

\begin{figure}[h!]
\begin{center}
\includegraphics[width=0.8\columnwidth,keepaspectratio=true]{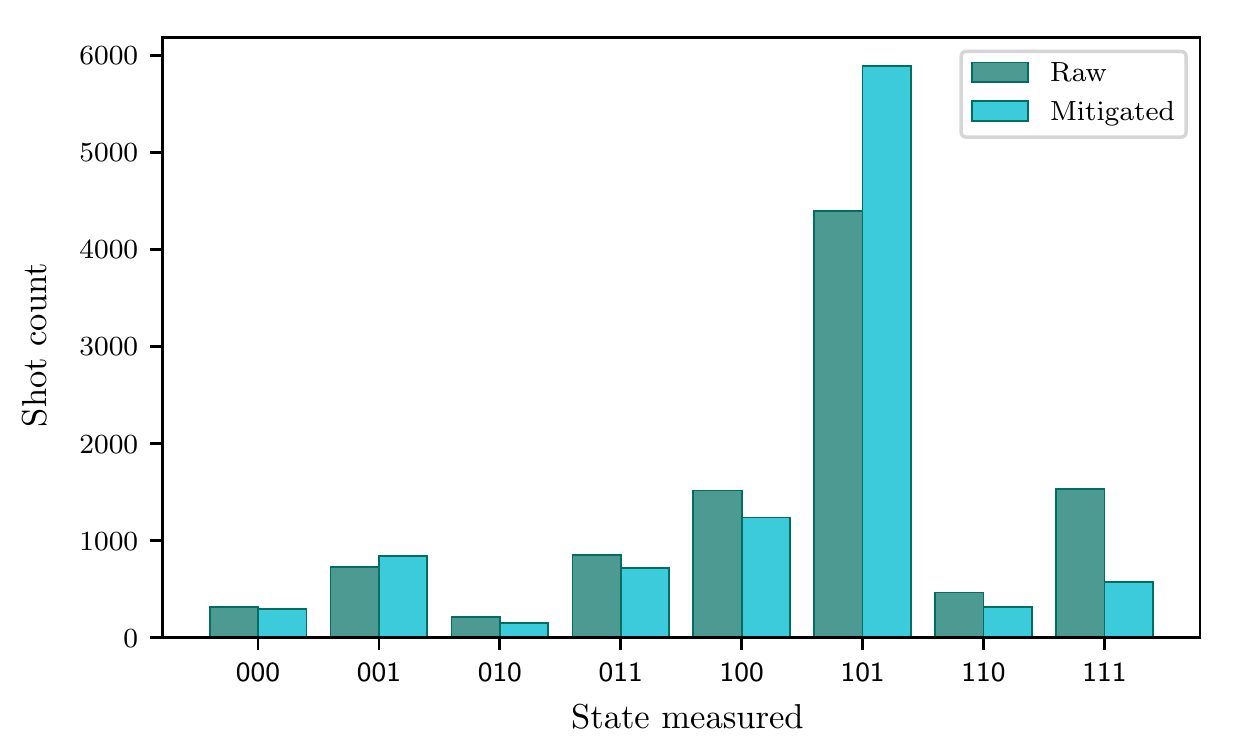}
\end{center}
\caption{\label{fig_qpe} Results of quantum phase estimation performed on Rigetti's Aspen-11 QPU, on a F$_2$ (2,2) example. QPE was performed for 3 bits of accuracy. The qubit Hamiltonian was obtained by the Bravyi-Kitaev transformation, and reduced to a single-qubit Hamiltonian using symmetries. First-order Trotterisation was used with two Trotter steps. The raw results are those directly obtained from the QPU, while mitigated results were obtained using a basic form of measurement error mitigation.}
\end{figure}

We study a simple example, taking the F$_2$ molecule and considering a (2,2) active space consisting of the canonical $\sigma$ and $\sigma^*$ bonding and anti-bonding orbitals. The fermionic Hamiltonian is mapped to a qubit Hamiltonian using the Bravyi-Kitaev transformation.\cite{Bravyi2002, Seeley2012} Because there are four spin-orbitals, we would typically expect the qubit Hamiltonian to act on four qubits. However, in this simple case, this can be reduced to a single-qubit Hamiltonian by applying particle number, spin, and spatial symmetries. In particular, the final Hamiltonian can be written
\begin{equation}
    \hat{H} = c_0 + c_1 \hat{Z} + c_2 \hat{X}
\end{equation}
where $\hat{Z}$ and $\hat{X}$ are Pauli-Z and Pauli-X operators, and $c_0 = -198.302013$ Ha, $c_1 = 0.464213$ Ha and $c_2 = 0.210370$ Ha. This reduction makes the problem of performing QPE tractable on current QPUs, provided that further parameters are chosen appropriately. A similar QPE experiment has recently been performed on a neutral atom quantum computer, although was performed on a different example system, H$_2$\cite{ColdQuanta}

We perform QPE using $\hat{U} = e^{-i\hat{H}t}$ as the evolution operator. Here, $t$ is chosen so that the eigenvalues of $Ht$ lie between $-\pi$ and $+\pi$. The operator $U$ is approximated using first-order Trotterisation with two Trotter steps. Lastly, we perform QPE to 3 bits of precision using the textbook version\cite{NielsenChuang} of the QPE algorithm. We therefore use $3$ ancilla qubits, together with a single data qubit on which the Trotterised $\hat{U}$ operator is applied. Thus, the experiment requires $4$ qubits total. The data qubit is initialised from the Hartree--Fock determinant, which ensures a strong overlap with the ground-state wave function.

We compiled the QPE circuit to native gates of the Aspen-11 QPU several times, and tested $9$ such compiled circuits on the QPU. This included testing different sets of qubits. We present results from the compiled circuit which resulted in the strongest signal. However, note that this is likely far from optimal for the QPU used, and we have not performed extensive tests to optimize the compiled circuit. Results are presented in Fig.~\ref{fig_qpe}. The circuit was performed for $10^4$ shots. A basic error mitigation technique was applied which aims to remove readout error. This is error that occurs in the quantum computation during measurement operations. The approach used is the ``unbiased'' technique described by Bravyi \emph{et al.}~\cite{Bravyi2021} We performed this experiment four times in total for the same native circuit, in order to demonstrate that a strong signal can be reproduced. The results of the additional repeated experiments are presented in the Supplementary Material.

A strong signal of around $59\%$ is seen on the state $| 101 \rangle$, which indicates an energy of $0.101$ in binary, or $0.625$ in decimal. This compares to a signal of $90\%$ on a noiseless simulator. After including appropriate shifting and factors from the Hamiltonian, this QPE result corresponds a final energy estimate of $-198.8080$ Ha, compared to an exact result (within the active space) of $-198.8117$ Ha. Thus, the error is only $3.7$ mHa. Given that only 3 bits of precision are obtained, this accuracy is somewhat fortuitous; the exact phase estimation result for the Trotterised $\hat{U}$ used is 0.636277, which lies very close to 0.101 in binary.

In general, achieving chemical accuracy with QPE will require many more bits of precision, and more Trotter steps in the representation of $\hat{U} = e^{-i\hat{H}t}$. The required circuit quickly becomes extremely deep, requiring QPU error rates which are not achievable, even in the long term. Performing these deeper QPE circuits will only be possible on fault-tolerant quantum computers which implement quantum error correction. Even then, current resource estimations suggest that QPE will become expensive for very large active spaces\cite{Reiher2017, Lee2021}. Therefore, embedding methods will be of crucial importance.

\section*{\sffamily \Large The Bond Dissociation Problem}
 
Bond dissociation is a problem that single reference methods struggle to describe correctly. \cite{knowles2000ab} To begin with, the spin restricted Hartree-Fock model characterizes a pair of electrons with the same spatial orbital regardless of the separation between nuclei and hence it describes ionic bond dissociation rather than the energetically more favorable homolytic process.  Spin unrestricted solutions provide a qualitatively correct solution in many cases, but the resulting state is no longer a pure spin-eigenstate. Typically, a small active space multiconfigurational calculation is the best way to ensure a qualitatively correct solution that is also a spin-eigenstate.  Such a multiconfigurational state can then be used as a reference for further correlation treatment. For diatomic molecules, a small active space is usually sufficient for generating such a state, but the necessary active spaces may increase with the complexity of the molecule. While the promise of quantum computers is to accelerate large active space calculations, bond dissociation problems may serve as prototypical applications, all the more so since the selection of active spaces is relatively easy.

\begin{figure}[h!]
\begin{center}
\includegraphics[width=1\columnwidth,keepaspectratio=true]{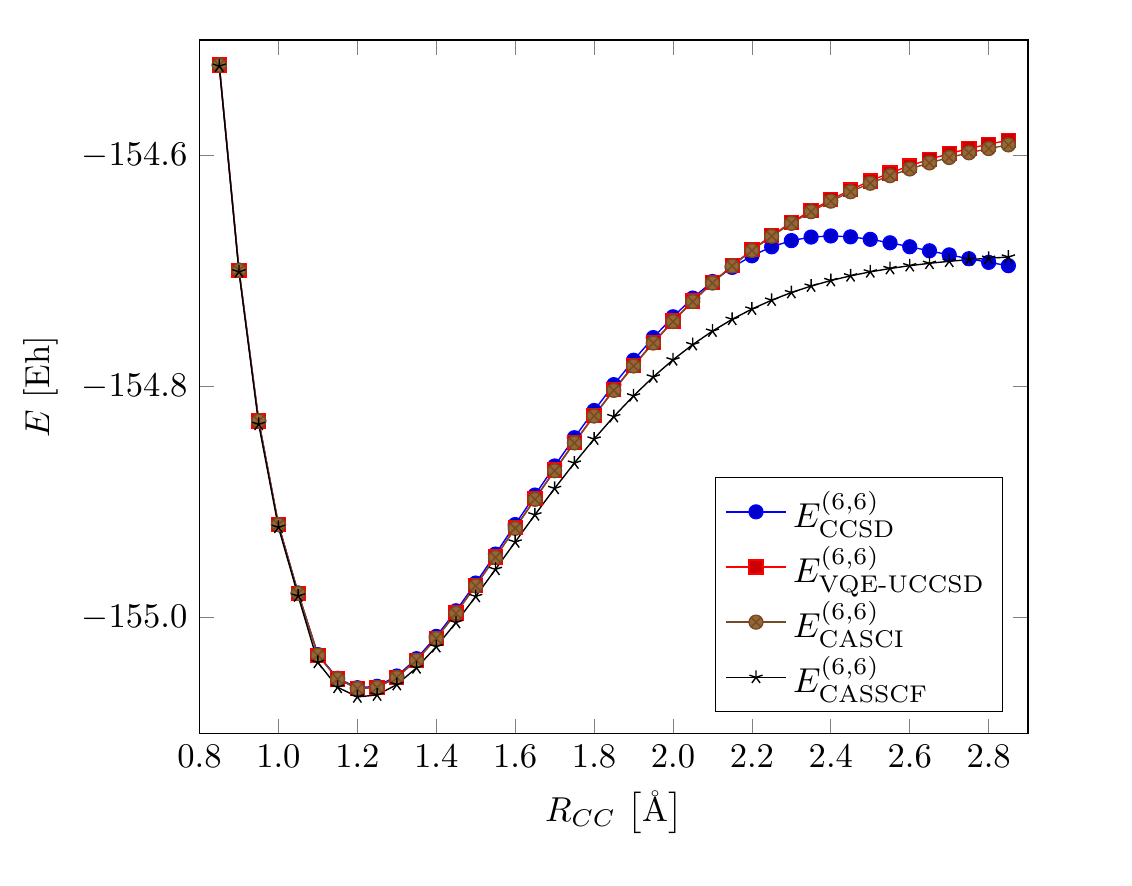}
\end{center}
\caption{\label{fig1} Bond dissociation curve of dimethyl acetylene at various levels of theory within the same (6,6) active space.}
\end{figure}

Coupled cluster methods also have difficulty in describing dissociation curves, especially for multiple bonds at larger distances. \cite{taube2006new} Thus, investigating the dissociation curve of $\text{N}_2$ has become a standard test for multireference CC methods. Since our goal is eventually to use multiscale embedding methods, $\text{N}_2$ is too small a test system. Instead, dimethyl acetylene will be considered, which contains a triple bond and two methyl groups that will serve as the environment. Choosing the two carbon atoms associated with the triple bond as the central fragment yields the three C$-$C bonds as the occupied orbitals in the active space. Three virtual orbitals are then selected based on the perturbative criteria described earlier. The Supplementary Material contains the HF, MP2, CASCI, CASSCF, CCSD and CCSD(T) dissociation curves for this system, while Fig.~\ref{fig1} shows the CCSD, VQE-UCCSD, CASCI/QPE and CASSCF results in the chosen (6,6) active space. In Fig.~\ref{fig1}, the CASSCF curve is the closest to the FCI solution of the entire system, while the other curves are slightly above it around the equilibrium bond length ($\sim$1.2 \AA). The CCSD curve in the (6,6) active space exhibits the well-known problem of CC bond dissociation curves: it drops below the exact solution and approaches smaller values as the bond distance increases. The MP2, CCSD and CCSD(T) methods within the full space yield similar results, and the problem does not vanish at higher orders of CC theory, except if the CC expansion itself becomes equivalent to FCI for a given system. In CC theory, this behavior is due to the non-Hermitian nature of the CC Hamiltonian and to the fact that the CC problem is not solved variationally. Variational CC methods do not have this problem. Since the VQE-UCCSD approach used in quantum computing is a variational approach, the bond dissociation curve does not have the pathological behaviour associated with traditional CCSD. The same is true of CASCI that can be regarded as the classical computing equivalent of QPE. The VQE and QPE/CASCI curves lie very close to each other, with the QPE one being slightly lower. They are both well above the CASSCF curve, since the latter also contains orbital relaxation effects.  Note that it is also possible to approach the CASSCF curve using a hybrid quantum-classical computing approach in which the CI problem is solved on the quantum computer and the orbital relaxation steps are carried out using classical computational approaches.\cite{tilly2021reduced}

\begin{figure}[h!]
\begin{center}
\includegraphics[width=1\columnwidth,keepaspectratio=true]{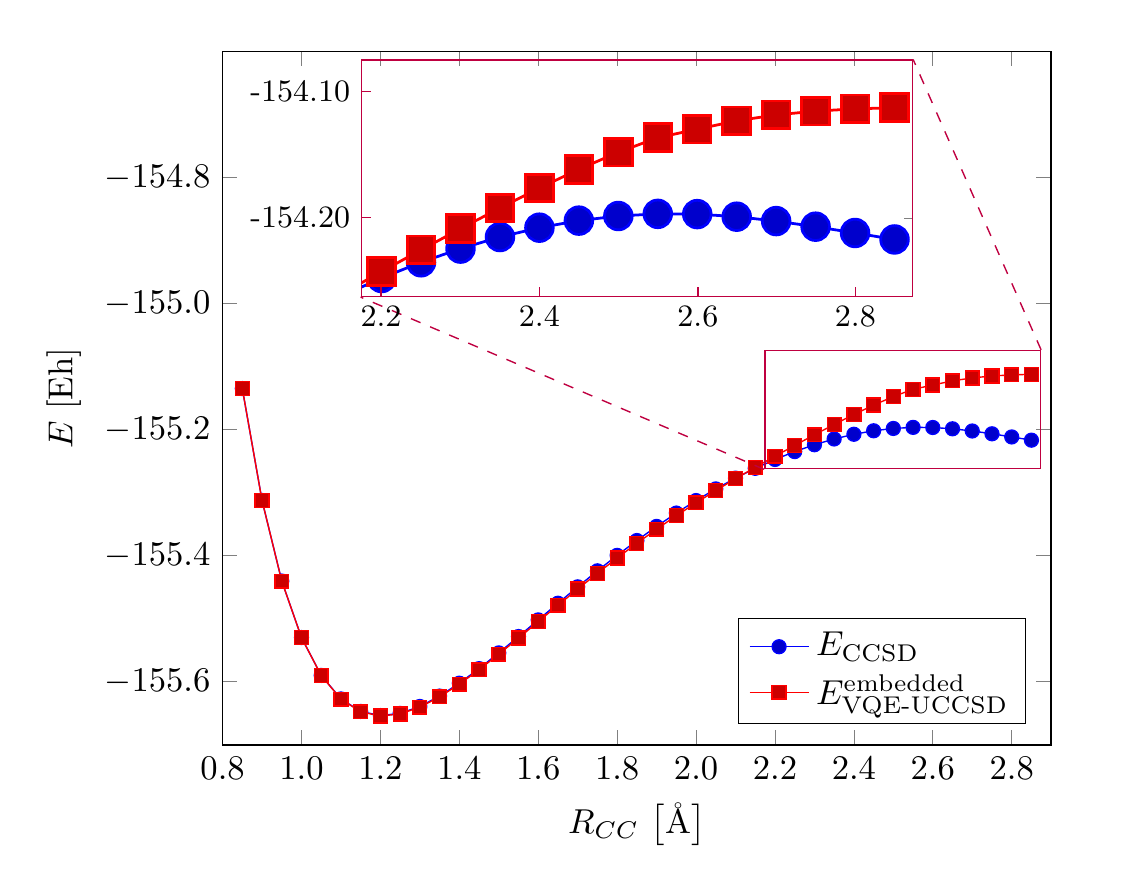}
\end{center}
\caption{\label{fig2} The CCSD and the embedded VQE-UCCSD dissociation curves of dimethyl acetylene.}
\end{figure}

Next, let us consider embedding approaches. In analogy with Eq.~\eqref{EMBsub}, an embedded VQE approach can be defined as
\begin{equation}
E^{\text{embedded}}_{\text{VQE-UCCSD}} = 
E_{\text{CCSD}} + E^{(6,6)}_{\text{VQE-UCCSD}} - E^{(6,6)}_{\text{CCSD}}.
\label{Eemb}
\end{equation}
Assuming that the pathological behaviour of the CCSD curve is mostly due to contributions from the active space, this formula attempts to correct the CCSD curve by difference between the VQE-UCCSD and CCSD curves calculated within the (6,6) active space. At the same time, this approach retains a high-level CCSD treatment for the environment. The result is plotted in Fig.~\ref{fig2} along with the uncorrected CCSD curve. The most important conclusion is that the embedded curve shows the qualitatively correct dissociation behaviour, although the dissociation energy itself is still too high. A further source of error is the fact that the amplitudes in the various terms of the subtractive embedding formula are not optimized together. The consequences of this are most evident if the first and second derivatives are also calculated from finite differences (see Supplementary Material). The embedded energy curve approximates the CASSCF curve well apart from a shift, and both show behaviour expected from the Morse model potential. However, while the CASSCF derivative curves are Morse-like, the embedded curves have humps at large bond length values. It is important to note however that the zero values of the gradient curve are reproduced by the embedded curve, which is the crucial feature as far as geometry optimization is concerned. Even the second derivatives have the expected signs at large distances. Thus, the embedded VQE-UCCSD (or QPE) approach should be suitable for studying bond dissociation problems. In most practical applications, there is a further mitigating circumstance: bond dissociation is usually followed by the formation of a new bond. This is a process that is characterized by a double well potential in which the extreme long distance part of the bond dissociation curves is replaced by another potential well, where embedding methods perform much better.

\section*{\sffamily \Large Enzyme Catalysis}

Enzymes are biological catalysts that play an important role in a number of reactions which mostly involve the breaking and formation of chemical bonds. As classical force fields have difficulty in treating the latter accurately, QM/MM methods are typically used to characterize such reactions and as some of these involve strong correlation, quantum computing may be especially useful in this context. A quantum computational study of protein-ligand interactions using DMET has already been carried out recently.\cite{kirsopp2021quantum} In the following, we will illustrate how an active space can be selected for a specific enzyme reaction and use the embedding treatment developed above for the environment.  The smallest meaningful active space will be selected, so that the calculation will actually reflect what is possible or nearly possible on today's quantum computers. However, the principles remain the same for larger active spaces and the procedure can be easily adapted to the quantum computers of the future.

\begin{figure}[h!]
\begin{center}
\includegraphics[width=1\columnwidth,keepaspectratio=true]{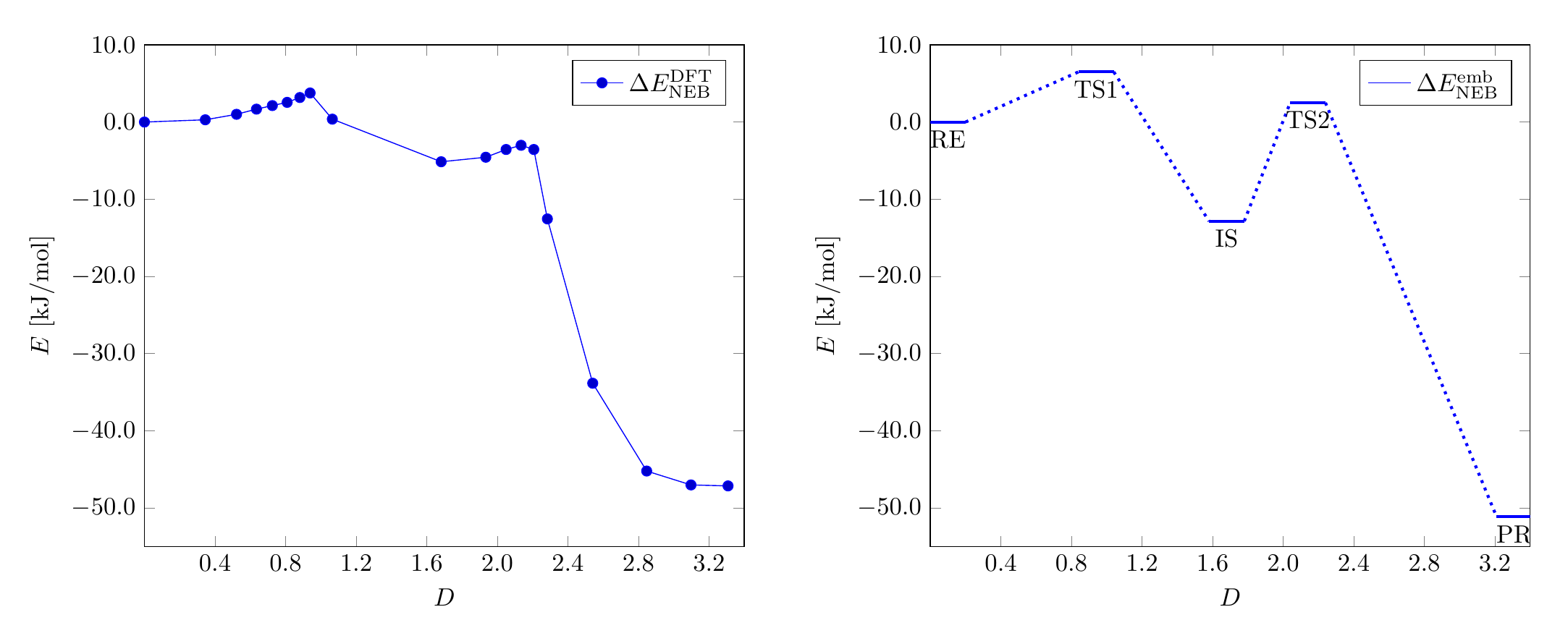}
\end{center}
\caption{\label{fig3} Left: minimum energy pathway of $H_2$ activation process calculated at the TPSS (DFT) level. Right: the extrema calculated at the DFT geometries using the embedding method.}
\end{figure}

\begin{table}
\begin{tabular}{lcccccc}
\hline
\textbf{Species} & $E_{\text{CCSD}}$ & $E_{\text{VQE-UCCSD}}^{(4,4)}$ & $E_{\text{CCSD}}^{(4,4)}$ & $\Delta E^{\text{DFT}}_{\text{NEB}}$ & $\Delta E_{\text{VQE-UCCSD}}^{(4,4)}$ & $\Delta E^{\text{emb}}_{\text{NEB}}$ \\ 
\hline
RE & -4214.326840 & -4193.943270 & -4193.943271 & 0.00 & 0.00 & 0.00 \\
TS1 & -4214.324367 & -4193.934798 & -4193.934797 & 3.77 & 22.24 & 6.49 \\
IS & -4214.331724 & -4193.945170 & -4193.945171 & -5.15 & -4.99 & -12.82 \\
TS2 & -4214.325888 & -4193.936151 & -4193.936144 & -3.01 & 18.69 & 2.48 \\
PR & -4214.346323 & -4193.975599 & -4193.975600 & -47.15 & -84.88 & -51.15 \\
\hline
\end{tabular}
\caption{\label{tbl1} The various components of the embedding method for the [Fe] hydrogenase model system in atomic units and the embedded energy differences ($\Delta E^{\text{emb}}_{\text{NEB}}$) in kJ/mol.}
\end{table}

As an example, we will consider the hydrogen activation mechanism catalyzed by [Fe] hydrogenase investigated in detail elsewhere.\cite{finkelmann2014hydrogen} The reaction can be briefly summarized as
\begin{equation}
\ce{methenyl-H4MPT+ + H2 <=> methylene-H4MPT + H+}
\end{equation}
One possible mechanism involves the concerted cleavage of $\ce{H2}$ in which the resulting $\ce{H+}$ is transferred to an oxypyridine ligand. Although experimentally it is unclear whether the cleavage is concerted or consecutive,\cite{finkelmann2014hydrogen, huang2019atomic} we will consider the two step mechanism, since it is somewhat more complicated and more illustrative as a proof of principle for our method. In the first step, the $\ce{H2}$ molecule is bound to the iron core of [Fe] hydrogenase and reacts with the oxygen of oxypyridine
\begin{equation}
\ce{Fe(II)\bond{...}H2 + OR^- -> Fe(II)\bond{...}H- + HOR}.
\end{equation}
In the second step, the hydride intermediate reacts with $\ce{methenyl-H4MPT+}$
\begin{equation}
\ce{Fe(II)\bond{...}H- + methenyl-H4MPT+ -> Fe(II) + methylene-H4MPT}.
\end{equation}
Given these reaction steps, the smallest fragment that can be chosen consists of the two H atoms of $\ce{H2}$, the Fe atom of the core, the O of oxypyridine and a C of $\ce{methenyl-H4MPT+}$. The procedure described above then yields eight candidates for active occupied orbitals: the five ``3d'' orbitals of Fe, the two lone pairs of O and the $\ce{H2}$ bond orbital. Of these, the most important are the lone pair of O that points towards $\ce{H2}$, as this will become the $\ce{HO}$ bond orbital, and the $\ce{H2}$ bond orbital itself, which becomes a doubly occupied orbital in $\ce{H-}$ and, eventually, the $\ce{CH}$ bond in $\ce{methenyl-H4MPT+}$.  We will thus select the latter two as active and let another two virtual orbitals be selected automatically to yield a (4,4) active space. 

\begin{figure}[h!]
\begin{center}
\includegraphics[width=1\columnwidth,keepaspectratio=true]{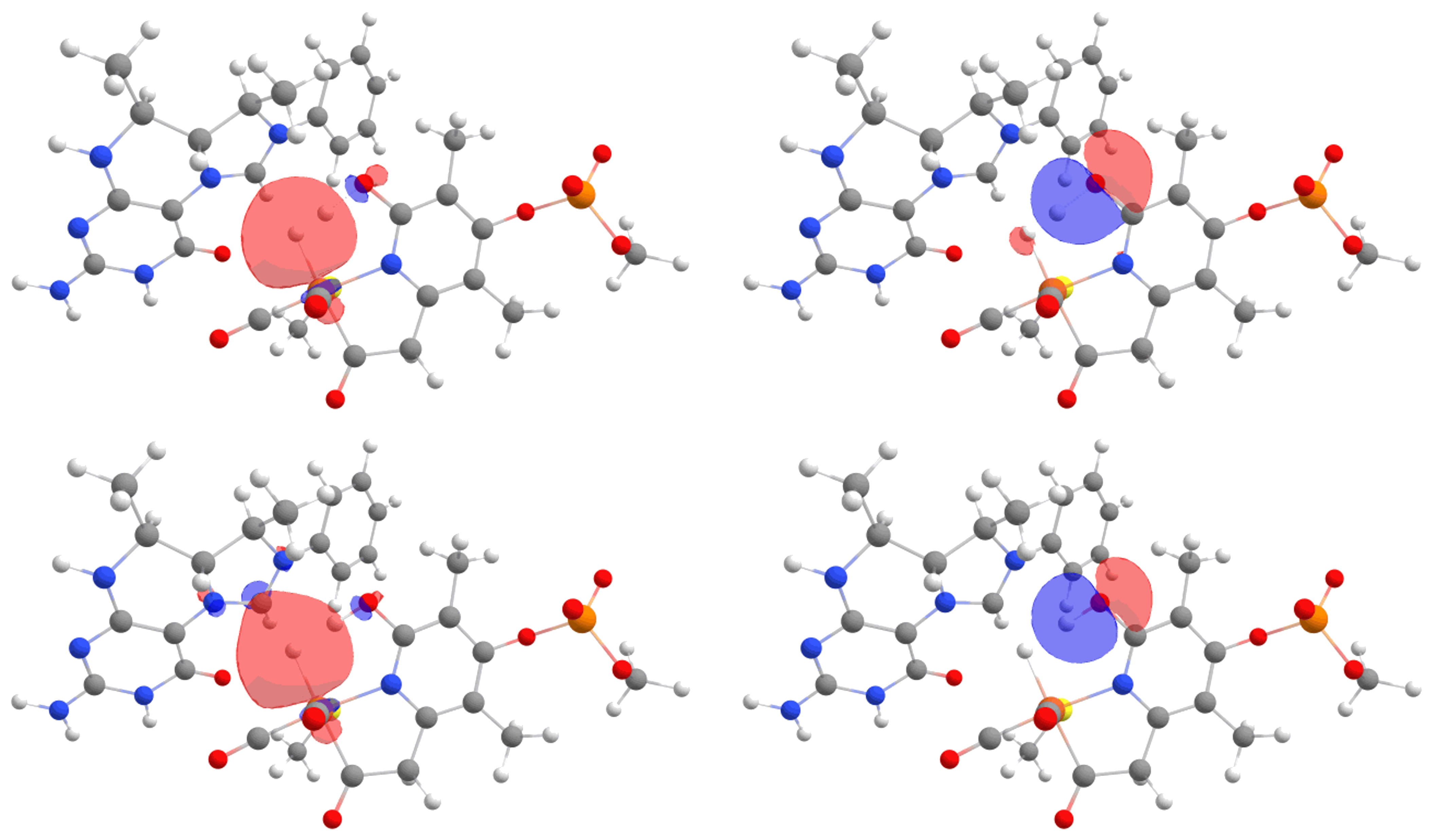}
\end{center}
\caption{\label{fig4} Top row: the active occupied orbitals at the first transition state geometry. Bottom row: the active occupied orbitals at the second transition state geometry.}
\end{figure}

We begin our investigation by finding a minimum energy pathway (MEP) for the two step reaction above using the NEB (Nudged Elastic Band) approach at the DFT level, see Fig.~\ref{fig3}. The structures with the locally highest energy can be used to find the two transition states, which are shown together with the two active occupied orbitals in Fig.~\ref{fig4}. This yields a total of five structures: the reactant (RE), the product (PR), and intermediate state (IS) and two transition states (TS1, TS2).  A formula closely related to Eq.~\eqref{Eemb} for the selected (4,4) active space gives the energetic profile using the embedding approach,
\begin{equation}
E^{\text{embedded}}_{\text{VQE-UCCSD}} = 
E_{\text{DLPNO-CCSD}} + E^{(4,4)}_{\text{VQE-UCCSD}} - E^{(4,4)}_{\text{CCSD}}.
\end{equation}
The results obtained from this formula and summarized in Table~\ref{tbl1} are significantly different from the DFT ones. Most importantly, the first (rate determining) transition state is $\sim$2.7 kJ/mol higher in the embedded calculation, a difference that corresponds to a factor of 5.6 in the rate constant. Another important difference is that the intermediate state is more stable when compared to the DFT prediction, which may explain why this structure is so hard to locate.\cite{finkelmann2014hydrogen} While the embedded energy profile is significantly different from the DFT result, the active space correction is in the order of a microhartree, same as the convergence threshold of the energy in these calculations. Although for a small active space, the correction might be small especially if orbital relaxation effects are ignored, in this special case it should also be mentioned that a single reference description is more adequate for heterolytic dissociation processes than for homolytic ones discussed in the previous section. Lastly, if the energy differences are calculated solely from the active space results, they deviate significantly from the embedded results: in the case of the rate determining step, the difference is as large as 30 kJ/mol. This indicates the importance of dynamical correlation and the proper treatment of the environment in an embedded theory. All these qualifications notwithstanding, we have demonstrated how the active space may be selected for a QM/MM calculation on enzymes. Quantum computers will especially be useful if the active space is too large. In the present case, keeping the eight occupied fragment orbitals mentioned before and adding another eight virtual ones would yield an active space of (16,16) which is about the largest active space that can be handled by classical computers with exact CASCI. Increasing the fragment further to include more close lying orbitals would soon push the calculation beyond the limit of classical feasibility.

\section*{\sffamily \Large Photodynamic Therapy Drugs}

Photodynamic therapy is a mode of treatment that relies on an excited state reaction of a photosensitive compound and that is used to cure various types of diseases, including cancer. A critical property that eligible drugs must satisfy is that they must absorb light within the therapeutic window (630-850 nm). Computational methods may help predicting the absorption properties of compounds, though this may be a quite complicated task. First, due to the size of biological systems, some sort of multilayer QM/MM approach is required to account for environmental effects. Second, progress in the development of accurate excited state methods lags somewhat behind that of ground state methods.\cite{izsak2020single} Finally, nuclear motion must be taken into account more carefully, all the more so because the spectral intensity depends on the overlap of the nuclear wavefunctions of the two states that are involved in the excitation process. Ideally, a molecular dynamics simulation should be run and the absorption profile be calculated at several snapshots. Lacking that, a harmonic correction may be calculated at the equilibrium geometry.  As our goal is to perform a proof of concept calculation here, nuclear effects will be neglected and only vertical excitation energies will be calculated using our embedding scheme. The excited state analogue of Eq.~\eqref{Eemb} can be obtained by finding higher lying roots of the CCSD and CASCI problems and applying the equation directly to excitation energies ($\omega$),
\begin{equation}
\omega^{\text{embedded}}_{\text{CASCI/QPE}} = 
\omega_{\text{STEOM-DLPNO-CCSD}} + \omega^{(4,4)}_{\text{CASCI/QPE}} - \omega^{(4,4)}_{\text{STEOM-CCSD}}.
\end{equation}
To obtain CC excited states, we will use the STEOM-DLPNO-CCSD method which has been applied with success to obtain the absorption spectra of molecules up to a 100 atoms. \cite{berraud-pache2020unveiling,lechner2021excited} The CASCI calculations were performed using quantum phase estimation, running on a quantum simulator; details were provided in the ``Computational Details'' section. Given the difficulties facing various approaches on the classical computer, calculating excitation energies on the quantum computer may be especially fruitful as also indicated by recent investigations on dyes. \cite{gocho2021enhancement} It should also be noted that another area where quantum computers are predicted to have a long term advantage is dynamic simulations\cite{yuan2021fault} which would be advantageous in the accurate prediction of spectra. The latter will not be considered here in any detail. 

\begin{figure}[h!]
\begin{center}
\includegraphics[width=0.5\columnwidth,keepaspectratio=true]{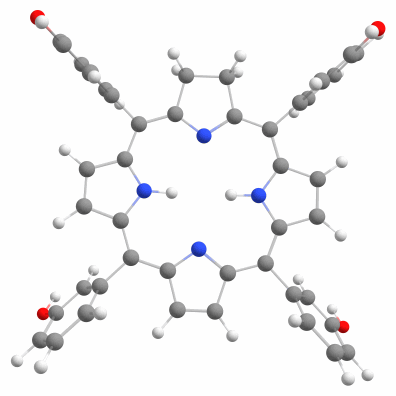}
\end{center}
\caption{\label{fig5} The optimized structure of the temoporfin molecule obtained from the last snapshot of a QM/MD run.}
\end{figure}

Temoporfin\cite{aslanoglu2020optical,devetta2018assessing} (Fig.~\ref{fig5}) has been chosen as a model system for further investigation. As a chlorin compound (partially hydrogenated porphyrin), temoporfin is known to absorb in the therapeutic window and forms the active component of the drug Foscan. Luckily, the low lying excitations are dominated by transitions between frontier orbitals. Denoting the HOMO$-1$, HOMO, LUMO, LUMO$+1$ orbitals as $h_1$, $h_0$, $l_0$, $l_1$, respectively, the first four STEOM states can be approximately written as
\begin{equation}
\Psi_1 \approx \frac{1}{\sqrt{2}}(\Phi^{l_1}_{h_1} - \Phi^{l_0}_{h_0}),
\end{equation}
\begin{equation}
\Psi_2 \approx \frac{1}{\sqrt{2}}(\Phi^{l_0}_{h_1} + \Phi^{l_1}_{h_0}),
\end{equation}
\begin{equation}
\Psi_3 \approx \frac{1}{\sqrt{2}}(\Phi^{l_1}_{h_1} + \Phi^{l_0}_{h_0}),
\end{equation}
\begin{equation}
\Psi_4 \approx \frac{1}{\sqrt{2}}(\Phi^{l_0}_{h_1} - \Phi^{l_1}_{h_0}).
\end{equation}
This means that forming a (4,4) active space from these canonical orbitals is a reasonable model for the excitation process. 

\begin{figure}
\begin{center}
\includegraphics[width=1\columnwidth,keepaspectratio=true]{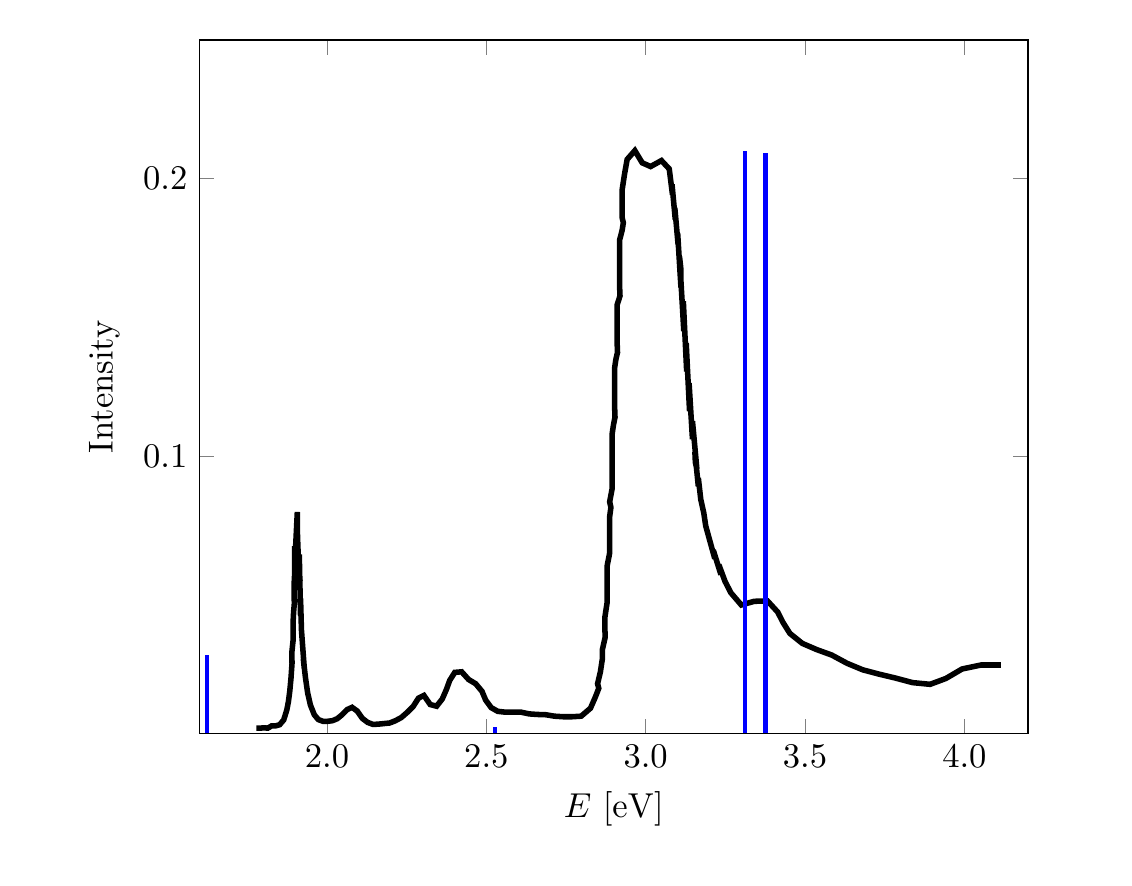}
\end{center}
\caption{\label{fig6} The experimentally measured absorption spectrum (black) and the embedded vertical excitation energies (blue).}
\end{figure}

\begin{table}
\begin{center}
\begin{tabular}{lccccc}
\hline
\textbf{State} & $E_{\text{STEOM}}$ & $E_{\text{CASCI/QPE}}^{(4,4)}$ & $E_{\text{STEOM}}^{(4,4)}$ & $\omega^{\text{embedded}}_{\text{CASCI/QPE}}$ & $f^{\text{osc}}_{\text{STEOM}}$\\ 
\hline
$\Psi_1$ & 1.634 & 3.318 & 3.331 & 1.621 & 0.3444 \\
$\Psi_2$ & 2.329 & 3.549 & 3.355 & 2.523 & 0.0295 \\
$\Psi_3$ & 3.428 & 5.367 & 5.486 & 3.309 & 2.5653 \\
$\Psi_4$ & 3.454 & 5.367 & 5.426 & 3.395 & 2.5545 \\
\hline
\end{tabular}
\end{center}
\caption{\label{tbl2} The various components of the embedding method for the temoporfin model system in atomic units and the embedded excitation energes ($\omega^{\text{embedded}}_{\text{CASCI/QPE}}$) in eV. The STEOM-CIS oscillator strength values ( $f^{\text{osc}}_{\text{STEOM}}$) are shown in atomic units. Note that $\Psi_3$ and $\Psi_4$ are degenerate in QPE to 10 bits of precision; this would be resolved with further bits measured.}
\end{table}

Fig.~\ref{fig6} and Table~\ref{tbl2} summarize the main findings of this section. The main features of the spectra are reproduced by our simple model, although with significant errors. First, note that only vertical excitation energies are reported here, without dynamic or harmonic corrections for the nuclei. The geometry used is an optimized geometry obtained from the last snapshot of an MD simulation, but it is certainly not a minimum structure but likely a higher order saddle point. STEOM vertical excitation energies are sensitive to the geometries used and also, when used with a harmonic correction, they typically require minimum structures on which the corresponding frequencies are calculated. For dynamics simulations, multiple snapshots would be required and performing the necessary STEOM calculations falls outside the scope of this study. The intensities are calculated from a simple CIS approximation\cite{ghosh2020new} and could also be improved in principle. As for the subsystem that could be treated on the quantum computer, a (4,4) active space is likely too small, as can be seen from the 1-2 eV difference when comparing either of the STEOM or CASCI values within the active space to the full STEOM calculation. The QPE results are within less that 0.01 eV agreement with results obtained from CASCI on a classical computer in the (4,4) active space, except for the highest lying root since it is degenerate with the one below in the QPE calculation at the given accuracy. Nevertheless, with these caveats in mind, it can be concluded that the embedded calculation reproduces the qualitative features of the absorption spectrum: the close lying high intensity peaks in the high energy region, and the two smaller intensity peaks in the low energy region. It is also noteworthy that the correction associated with a quantum computer is more substantial than it was in the [Fe] hydrogenase case. For the second excited state, it is as much as 0.2 eV, indicating the potential benefits of using quantum computers with larger active spaces for excited states to which the embedding methodology presented in this paper can be applied without significant modification.

\section*{\sffamily \Large CONCLUSIONS}

Our calculation on the F$_2$ model system using Rigetti's Aspen-11 QPU in combination with a basic error mitigation scheme demonstrates that quantum computers are well on the way to solving real chemical problems. Still, the application of quantum computers to the chemistry domain requires the vital step of selecting an active space that will be simulated in a fully quantum mechanical manner. This is the case for currently available quantum computers -- that are limited in the depth of the quantum circuits they can run -- and  will continue to be the case in the era of error-corrected quantum computers -- that will enable the full quantum simulation of much larger system sizes than it is possible to do with classical computational methods, albeit still smaller compared to the sizes required to simulate entire chemical systems of pharmaceutical interest.

To achieve the selection of the active space to be simulated, we have presented a scheme that combines the intuitive selection of the occupied space in terms of local orbitals on a fragment with the automatic selection of virtual orbitals using perturbation theory. We have demonstrated the usefulness of this procedure on a model system in which a triple bond is split. It has been established that our embedding approach improves on the weaknesses of traditional CCSD by correcting contributions within the active space, and it also allows for a high level treatment of the environment. Two applications that may serve as prototypes for tackling pharmaceutically relevant problems have also been presented. The case of [Fe] hydrogenase illustrates how our embedding approach can be applied to enzymatic processes that involve breaking and forming covalent bonds. While in this specific case the corrections from the simulated quantum computer are negligible, this is due to the specific properties of the reaction involved and it is expected that in general the quantum computer will significantly improve the description of strongly correlated systems with large active spaces. For the photochemical example of temoporfin, the corrections attributed to a simulated quantum computer are more significant. We did not aim for our results to be highly accurate here, but they do indicate a possible area of quantum advantage in the field of photochemistry, where classical methods struggle more than for the ground state. Both these calculations prove the central principle of our paper that our methodology can be applied to any problem of interest and we expect that as quantum computers will become more powerful, the same protocol can be used without any essential change in the algorithm. As active spaces increase, this will allow for the demonstration of practical quantum advantage in pharmaceutical applications.

\subsection*{\sffamily \large ACKNOWLEDGMENTS}

This work was performed as part of Astex's Sustaining Innovation Post-doctoral Program and it was supported via Innovate UK via the Quantum Commercialisation programme of the Industrial Strategy Challenge Fund (ISCF). We would also like to thank Gregor Giesen for technical help with ORCA and Georgi Stoychev for discussions on the NEB calculations.

\clearpage

\nocite{*}
\bibliography{embedding}% Produces the bibliography via BibTeX.

\end{document}

\maketitle

\clearpage

%\section*{\sffamily \Large THEORY}

\begin{figure}[h!]
\begin{center}
\includegraphics[width=0.8\columnwidth,keepaspectratio=true]{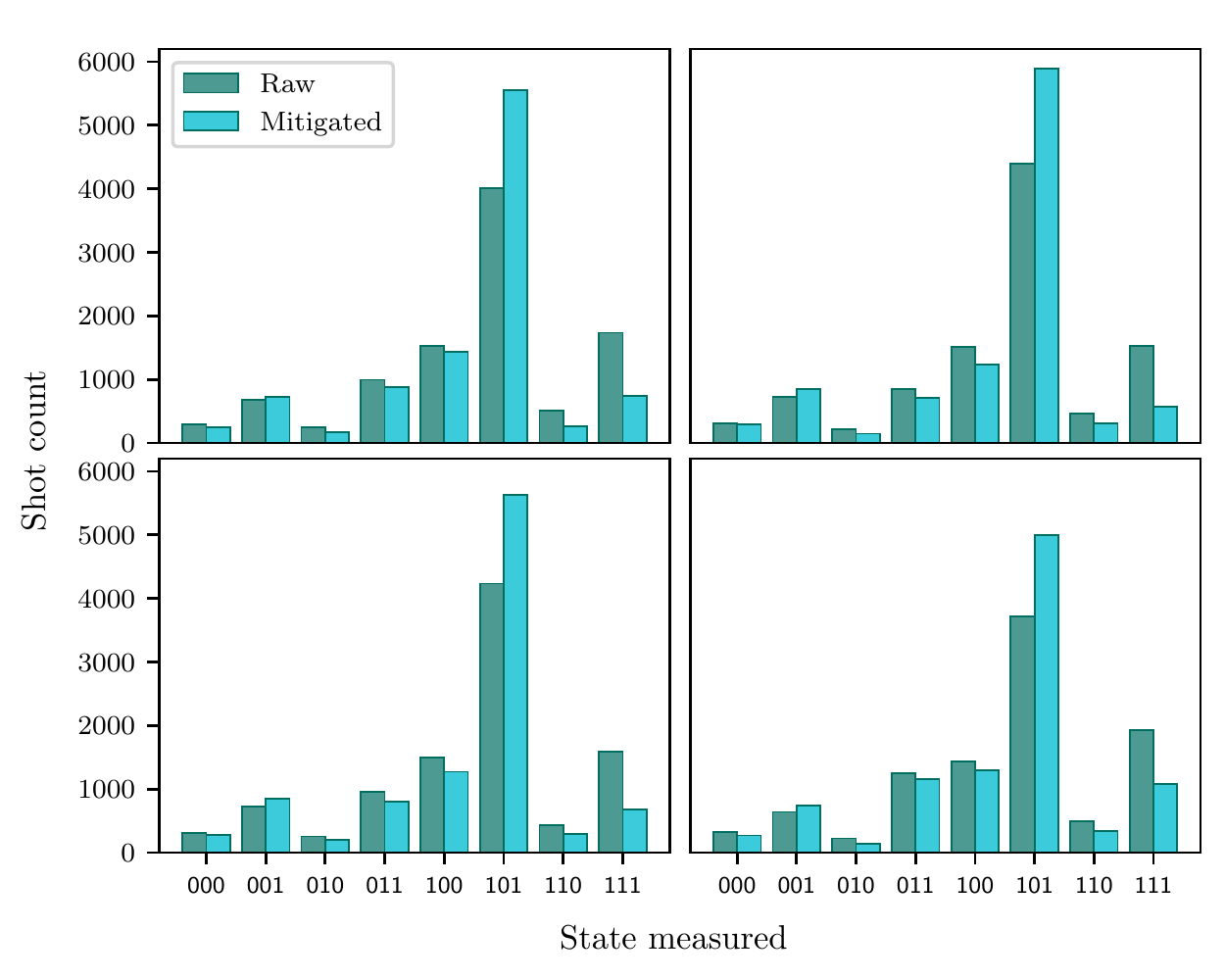}
\end{center}
\caption{\label{fig1} Results of quantum phase estimation performed on Rigetti's Aspen-11 QPU. The system is F$_2$ (2,2), and the qubit Hamiltonian is obtained using the Bravyi-Kitaev transformation, using symmetries to further reduce this to a single-qubit Hamiltonian. The four subplots show the results of four repeated experiment. Each experiment uses the same compiled circuit and is performed for 10,000 shots. Raw results are those directly measured on the QPU. Mitigated results were obtained by performing measurement error mitigation.}
\end{figure}

\clearpage

\begin{figure}[h!]
\begin{center}
\includegraphics[width=0.8\columnwidth,keepaspectratio=true]{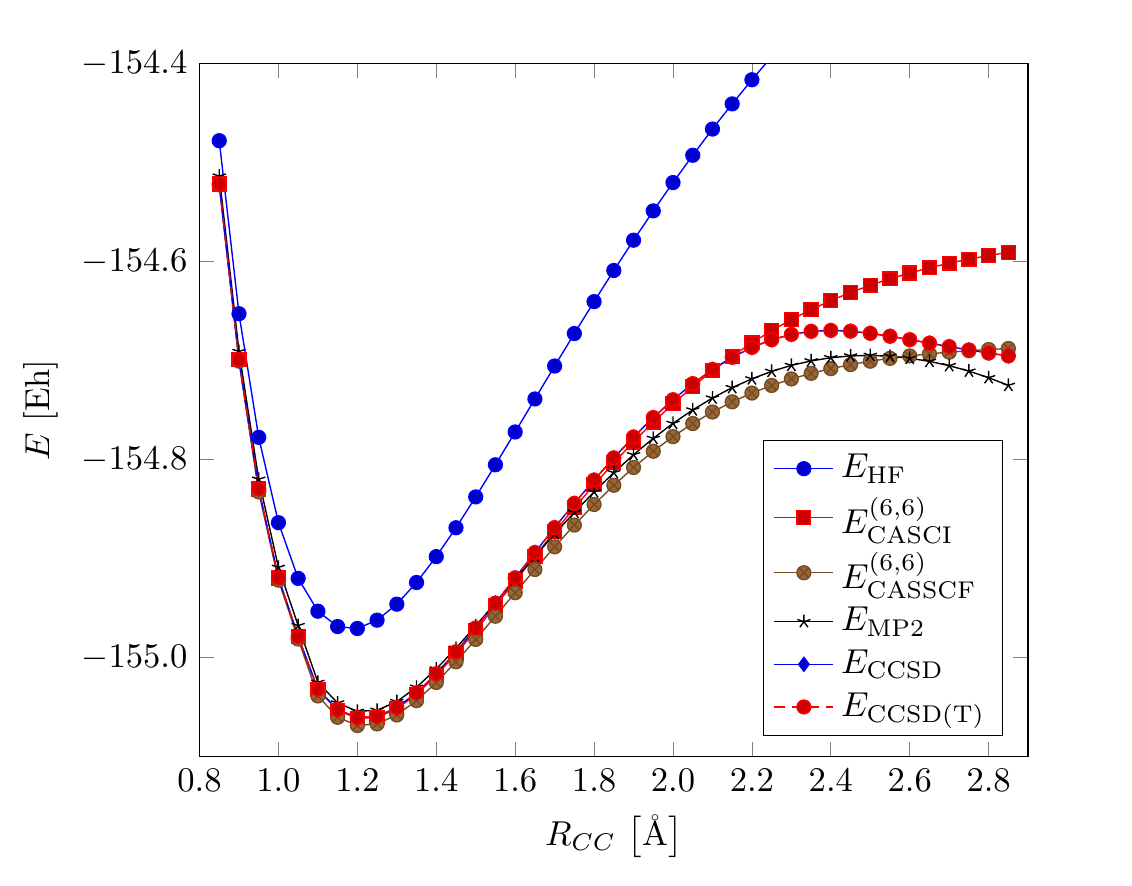}
\end{center}
\caption{\label{fig2} Bond dissociation curve of dimethyl acetylene at various levels of theory.}
\end{figure}

\clearpage

\begin{figure}[h!]
\begin{center}
\includegraphics[width=0.8\columnwidth,keepaspectratio=true]{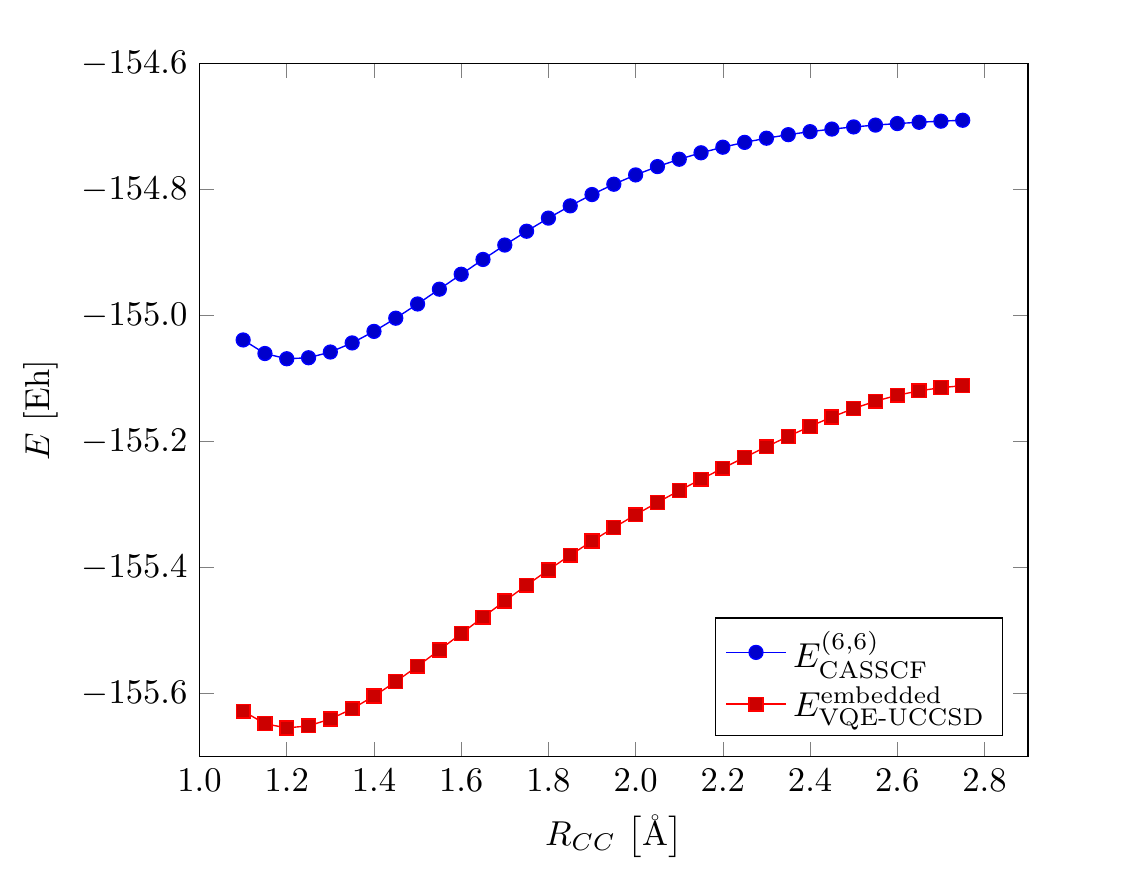}
\end{center}
\caption{\label{fig3} Bond dissociation curve of dimethyl acetylene at the CASSCF level within the (6,6) active space and at the embedded VQE-UCCSD level of theory.}
\end{figure}

\clearpage

\begin{figure}[h!]
\begin{center}
\includegraphics[width=0.8\columnwidth,keepaspectratio=true]{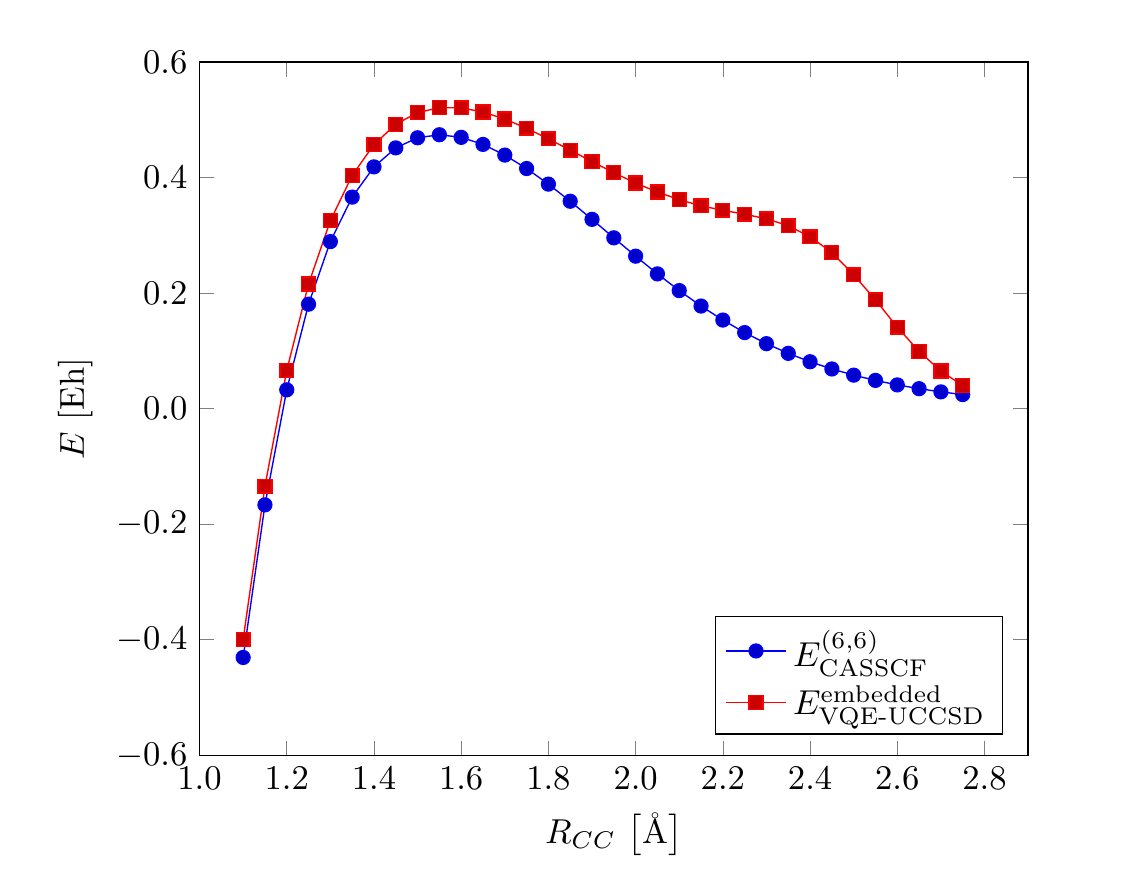}
\end{center}
\caption{\label{fig4} The first derivative from finite differences of the bond dissociation curve of dimethyl acetylene at the CASSCF level within the (6,6) active space and at the embedded VQE-UCCSD level of theory.}
\end{figure}

\clearpage

\begin{figure}[h!]
\begin{center}
\includegraphics[width=0.8\columnwidth,keepaspectratio=true]{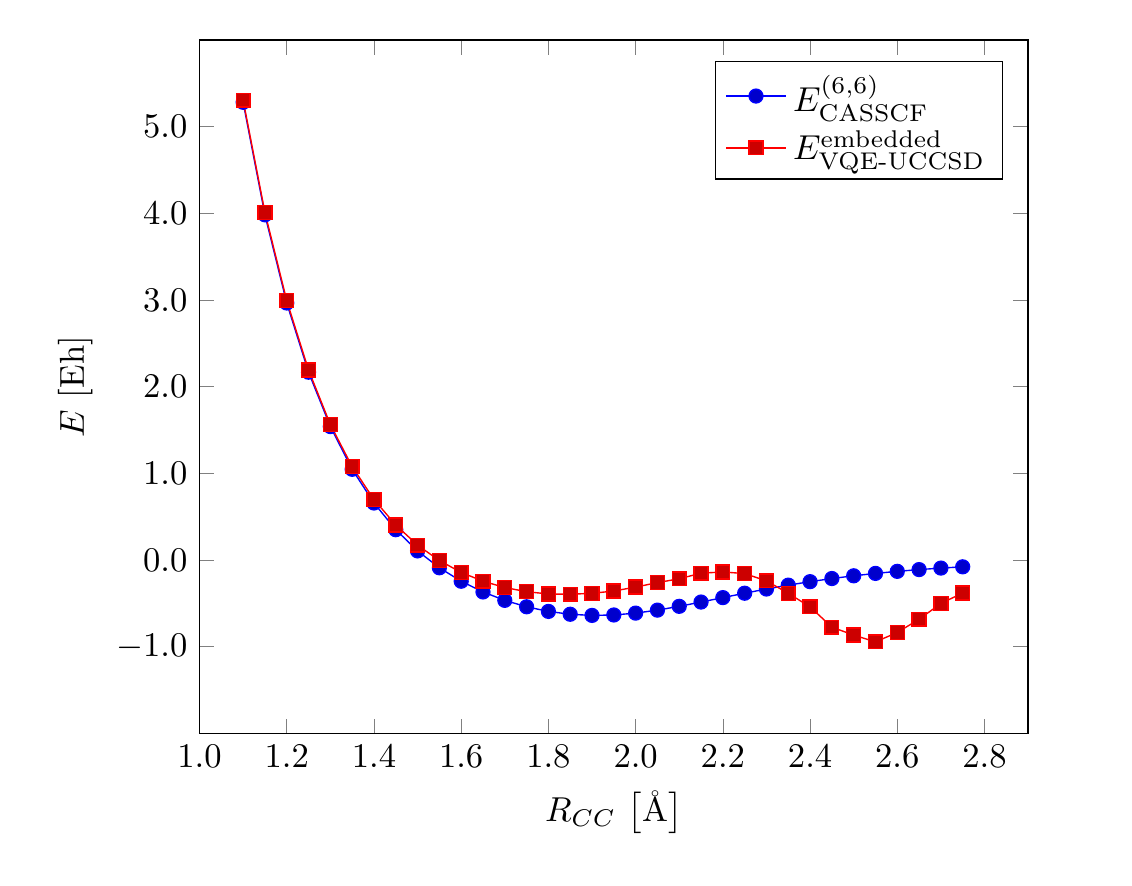}
\end{center}
\caption{\label{fig5} The second derivative from finite differences of the bond dissociation curve of dimethyl acetylene at the CASSCF level within the (6,6) active space and at the embedded VQE-UCCSD level of theory.}
\end{figure}